\documentclass[reprint,
amsmath,
amssymb,
floatfix,
superscriptaddress,
aps,
prx]{revtex4-2}
\usepackage{graphicx}
\usepackage{dcolumn}
\usepackage{verbatim}
\usepackage{bm}
\usepackage{color}

\usepackage{hyperref}%
\hypersetup{
    colorlinks=true,
    linkcolor=blue,
    citecolor=blue,   
   urlcolor=blue,
   }

\def\bea{\begin{eqnarray}}
\def\eea{\end{eqnarray}}
\def\beq{\begin{equation}}
\def\eeq{\end{equation}}

\def\f{\frac}
\def\p{\partial}

\def\la{\langle}
\def\ra{\rangle}

\def\qv{ {\bf q}}

\def\nv{ {\hat{n}}}
\def\nvperp{ {\hat{n}^{\perp}}}
\def\rv{ {\bf{r}}}

\def\fv{ {\bf{f}}}
\def\Fv{ {\bf{F}}}
\def\vv{ {\bf{v}}}
\def\uv{ {\bf u}}

\def\th{\theta}
\def\w{\omega}
\def\W{\Omega}
\def\a{\alpha}
\def\b{\beta}
\def\g{\gamma}
\def\d{\delta}
\def\l{\lambda}
\def\s{\sigma}
\def\taur{\tau_{\text{r}}}

\def\tauomega{\tau_{{\scriptscriptstyle \Omega}}}

\def\xiv{\boldsymbol{\xi}}
\DeclareMathOperator*{\Lim}{\underset{t \to \infty}{\lim}
}

\begin{document}
\title{Emergent Mesoscale Correlations in Active Solids with Noisy Chiral Dynamics}

\author{Amir Shee}%
\email[]{amir.shee@northwestern.edu}
\affiliation{Northwestern Institute on Complex Systems and ESAM, Northwestern University, Evanston, IL 60208, USA}
\author{Silke Henkes}%
\email[]{shenkes@lorentz.leidenuniv.nl}
\thanks{corresponding author}
\affiliation{Lorentz Institute for Theoretical Physics, LION, Leiden University, P.O. Box 9504, 2300 RA Leiden, The Netherlands}
\author{Cristi\'an Huepe}%
\email[]{cristian@northwestern.edu}
\thanks{corresponding author}
\affiliation{Northwestern Institute on Complex Systems and ESAM, Northwestern University, Evanston, IL 60208, USA}
\affiliation{School of Systems Science, Beijing Normal University, Beijing, People’s Republic of China}
\affiliation{CHuepe Labs, 2713 West August Blvd \#1, Chicago, IL 60622, USA}
\date{\today}%
%
%
\begin{abstract}
We present the linear response theory for an elastic solid composed of active Brownian particles with intrinsic individual chirality, deriving both a normal mode formulation and a continuum elastic formulation. 
Using this theory, we compute analytically the velocity correlations and energy spectra under different conditions, showing an excellent agreement with simulations.
We generate the corresponding phase diagram, identifying chiral and achiral disordered regimes (for high chirality or noise levels), as well as chiral and achiral states with mesoscopic-range order (for low chirality and noise). The chiral ordered states display mesoscopic spatial correlations and oscillating time correlations, but no wave propagation.
In the high chirality regime, we find a peak in the elastic energy spectrum that leads to a non-monotonic behavior with increasing noise strength that is consistent with the emergence of the `hammering' state recently identified in chiral glasses.
Finally, we show numerically that our theory, despite its linear response nature, can be applied beyond the idealized homogeneous solid assumed in our derivations.
Indeed, by increasing the level of activity, we show that it remains a good approximation of the system dynamics until just below the melting transition.
In addition, we show that there is still an excellent agreement between our analytical results and simulations when we extend our results to heterogeneous solids composed of mixtures of active particles with different intrinsic chirality and noise levels.
%
The derived linear response theory is therefore robust and applicable to a broad range of real-world active systems.
Our work provides a thorough analytical and numerical description of the emergent states in a densely packed system of chiral self-propelled Brownian disks, thus allowing a detailed understanding of the phases and dynamics identified in a minimal chiral active system.
\end{abstract}

\maketitle

\section{Introduction}
\label{sec_I}

Chirality is a fundamental property of most chemical, physical, and biological systems that is expected to naturally occur in active matter.
Indeed, in the context of self-propelled particles, chiral motion has been shown to spontaneously arise due to asymmetries in the self-propulsion forces or in the particle geometry~\cite{Kummel2013, Bechinger2016, Mano2017, Zhang2022}, or as a result of interactions with external fields~\cite{Grzybowski2000, Cruz2024}.

The relationship between activity and chirality has been considered in multiple contexts. Chiral motion has been observed experimentally in active biomolecules ~\cite{Jennings1901} such as proteins~\cite{Loose2014}, in microtubules~\cite{Sumino2012}, and in single cells, including bacteria~\cite{Brokaw1982, DiLuzio2005, Lauga2006} and sperm cells~\cite{Riedel2005, Nosrati2015}.
It has been studied theoretically for single circle swimmers~\cite{Van2008, Van-Teeffelen2009, Mijalkov2013, Volpe2014, Lowen2016, Sevilla2016, Morelly2019, Chepizhko2020, Caprini2023}, for the clockwise circular dynamics of {\em E. Coli}~\cite{Leonardo2011, Araujo2019}, and for circular and helical motion under chemical gradients~\cite{Bohmer2005, Taktikos2011}.
Groups of chiral swimmers have been shown to display unique forms of collective dynamics, whether chirality resulted from shape asymmetry~\cite{Gibbs2009, Gibbs2011, Denk2016, Bar2020, Zhang2020}, mass distribution~\cite{Campbell2017}, or catalysis coating~\cite{Archer2015}.
%
Chiral particles can also display a range of nonequilibrium phases, including a gas of spinners and aster-like vortices, rotating flocks with either polar or nematic alignment~\cite{Zhang2020}, and states displaying phase separation, swarming, or oscillations, among others~\cite{Lei2023}.

Several theoretical studies have shown that chirality can strongly affect the collective states that are typically found in achiral active systems.
In cases with explicit mutual alignment interactions (as in the Vicsek model), it has been shown that chiral polar swimmers display stronger flocking behavior than achiral ones, with higher levels of polarization in the ordered phase~\cite{Liebchen2017}, and that large rotating clusters with enhanced size and shape fluctuations can emerge~\cite{Levis2018}.
In cases with other types of angular interactions, a marked attenuation of motility-induced phase separation (MIPS)~\cite{Liao2018}, the emergence of vortex arrays~\cite{Kaiser2013}, and chirality-triggered oscillatory dynamic clustering~\cite{Liu2019} have been observed.
Chirality has also been found to affect the collective states of active particles without explicit alignment interactions.
For instance, chiral active Brownian particles can suppress conventional MIPS due to the formation of dynamical clusters that disrupt the MIPS clusters~\cite{Ma2022, Semwal2024}, and a quantitative field theory was developed to account for this suppression~\cite{Bickmann2022}.
%
Furthermore, chiral active particles with fast rotation have been found to form non-equilibrium hyperuniform fluids~\cite{Lei2019, Lei2019Science}.

Novel collective states have also been identified in inhomogeneous systems that combine different types of activity and chirality. 
For example, in a low-density environment, binary mixtures of passive and active chiral self-propelled particles exhibit transitions from mixed gels to rotating passive clusters, and then to homogeneous fluids~\cite{Hrishikesh2023}.
In addition, a mixture of active particles with different chirality frequencies can create complex combinations of clusters of different sizes, rotating at different rates~\cite{Levis2019PRE}.
Moreover, the combination of chiral and achiral swarming coupled oscillators leads to a range of novel behaviors, such as the formation of vortex lattices, pulsating clusters, or interacting phase waves~\cite{Ceron2023}.

Although most research on chiral systems has focused on liquid- and gas-like states, solid chiral active states have been found to naturally arise in systems such as groups of spinning magnetic particles \cite{bililign2022motile} or of starfish oocytes \cite{Tan2022} when hydrodynamic torque couplings are included, resulting in active chiral crystals~\cite{Drescher2009, Petroff2015, Yan2015, Huang2020SoftMatter, Ishikawa2020, Tan2022, Petroff2023}.
The interactions in these cases can be cast as nonreciprocal \emph{odd}-elastic viscous active couplings, to place them within the framework of odd active matter \cite{fruchart2023odd}.
The elastic coefficients then acquire non-symmetric contributions, and the resulting lack of energy conservation, as well as the polarity-position coupling, allow for wave propagation and work cycles.
%
Here we will consider a different class of systems, focusing on a minimal model of solid ``dry'' active matter, where chirality is introduced as part of the active forces, not as an active stress.
Since in this case there are no action-reaction effects in the active driving, activity cannot be recast as part of the stress tensor or in the elastic coefficients, and an odd-elasticity framework is not applicable.

Solid and dense active systems without chirality have received significant interest in recent years.
On one hand, the emergent states of self-propelled particles with self-alignment interactions have been studied in multiple contexts~\cite{Henkes2011, Lin2021, Baconnier2022, Lin2023, Baconnier2023, Xu2023}.
On the other hand, various dense and glassy active matter systems~\cite{Berthier2013, Berthier2014, Szamel2015, DebetsTJCP2023} without any alignment interaction have been described theoretically, using active Brownian particles in~\cite{Ni2013, Bi2016, Liluashvili2017, Nandi2018, Szamel2019, Mandal2020, Reichert2021, ReichertSoftMatter2021, ReichertEPJE2021, Debets2021, Debets2022, Janzen2022, Paoluzzi2022} and active Ornstein-Uhlenbeck particles in~\cite{Flenner2016, Szamel2016, Feng2017, Berthier2017, Flenner2020}. 
However, their chiral counterparts have so far received limited attention.
In one study, Debets et al.~\cite{Debets2023} examined the glassy dynamics of chiral active Brownian particles, showing that they exhibit highly nontrivial states and a non-monotonic behavior of the diffusion constant versus noise at high chirality that we also find in our system. 
In another study, Caprini et al.~\cite{caprini2024self} showed the emergence of rotating and oscillating states, deriving an analytical phase diagram by applying an active solid approach similar to that presented below, but in a different context.


In this paper, we investigate the collective dynamics of chiral active Brownian disks with elastic repulsive interactions at high densities, in the solid state. We identify and describe the different phases, finding an emergent mesoscopic length scale that can display or not oscillatory time correlations, depending on the ratio of chiral motion to rotational noise.
We derive an analytical active solid theory to describe these phases, using a normal mode approach and a continuum elasticity approach, both of which match our simulations.
In addition, we show that these results remain valid well into the nonlinear regime, just below the melting transition, and inform the dynamics of the fluid state. They also extend to different kinds of active binary mixtures, including mixtures of chiral and achiral particles, of chiral particles with different rotational speeds, and of chiral particles with different levels of rotational diffusion.

The paper is organized as follows. In Sec.~\ref{sec_II}, we describe our two-dimensional active solid model of densely packed self-propelled disks with elastic interactions and intrinsic individual chirality. 
In Sec.~\ref{sec_III}, we overview the phase space of dynamical regimes as a function of chirality and rotational diffusion. In Sec.~\ref{sec_IV}, we present our analytical results. We first calculate the orientation autocorrelation functions using a Fokker-Planck approach; then describe the normal mode formalism for active solids, calculating the average energy per mode and the spatial velocity correlations; and finally describe the continuum elastic formulation.
In Sec.~\ref{sec_V}, we characterize the dynamics described by our results and compare them to simulations.
In Sec.~\ref{sec_VI}, we examine the melting regime by increasing the level of activity. 
In Sec.~\ref{sec_VII}, we extend our results to heterogeneous mixtures of disks with different levels of activity and chirality. 
Finally, Sec.~\ref{sec_VIII} presents our conclusions.

\begin{figure*}[!ht]
\includegraphics[width=5.75cm]{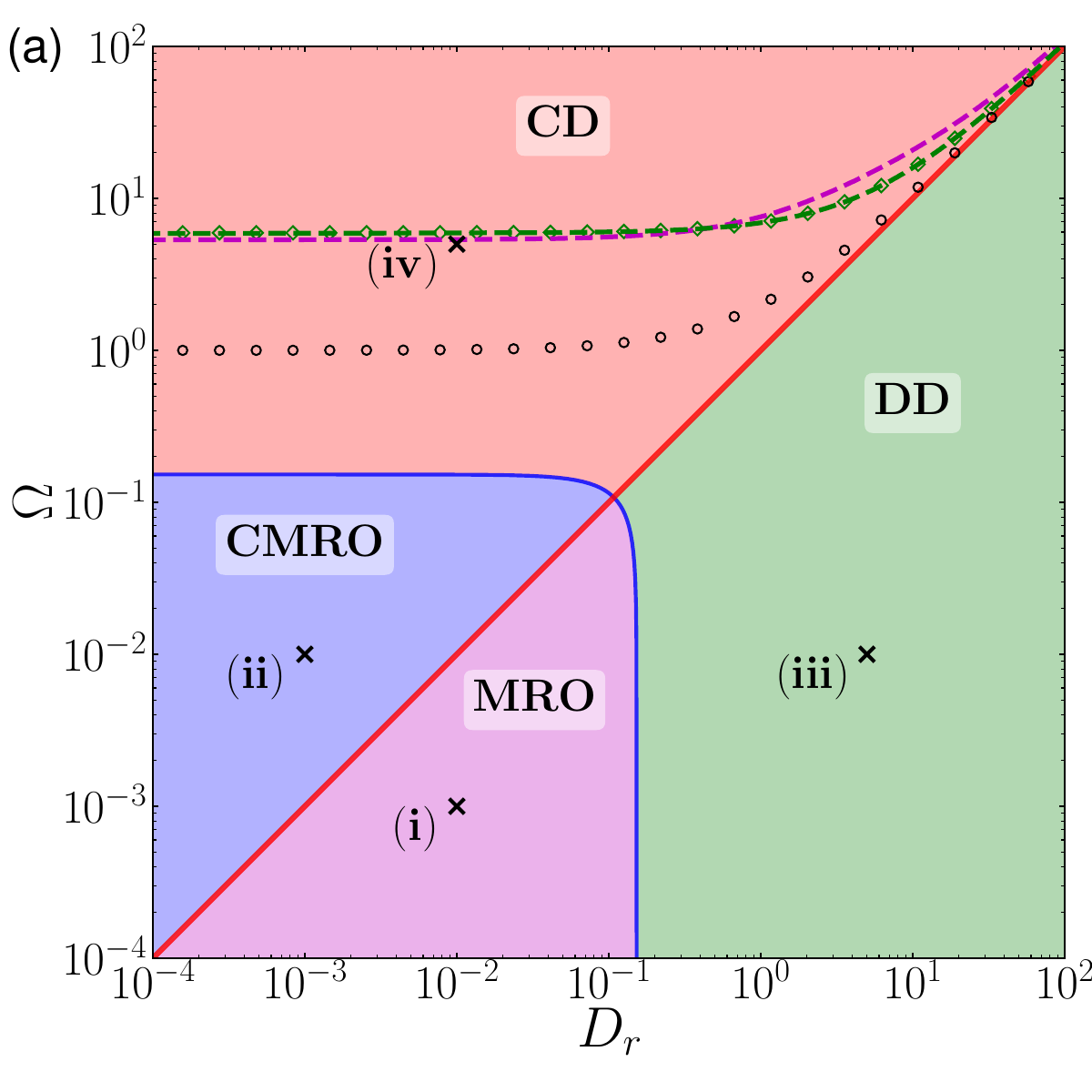}
\includegraphics[width=5.75cm]{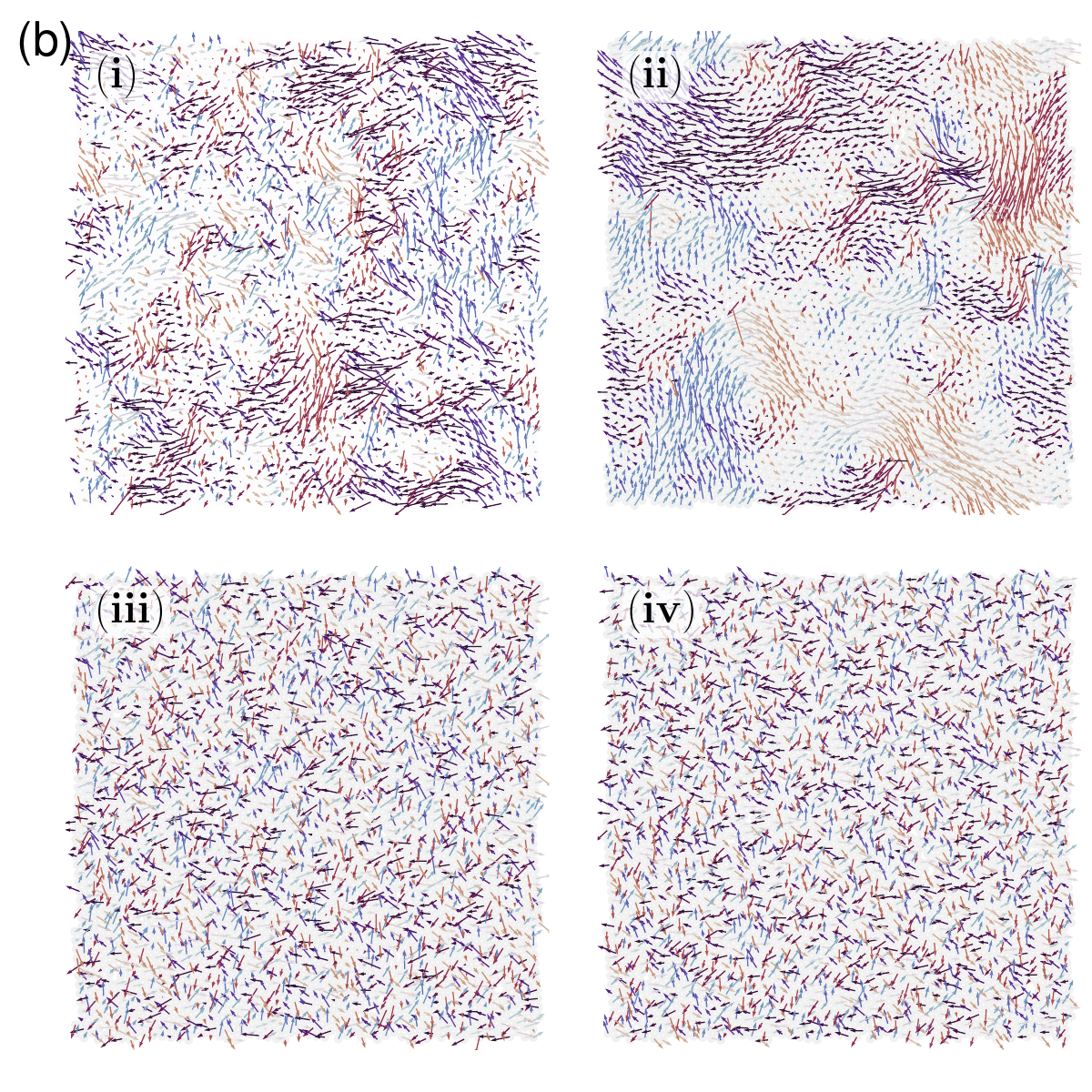}
\includegraphics[width=5.75cm]{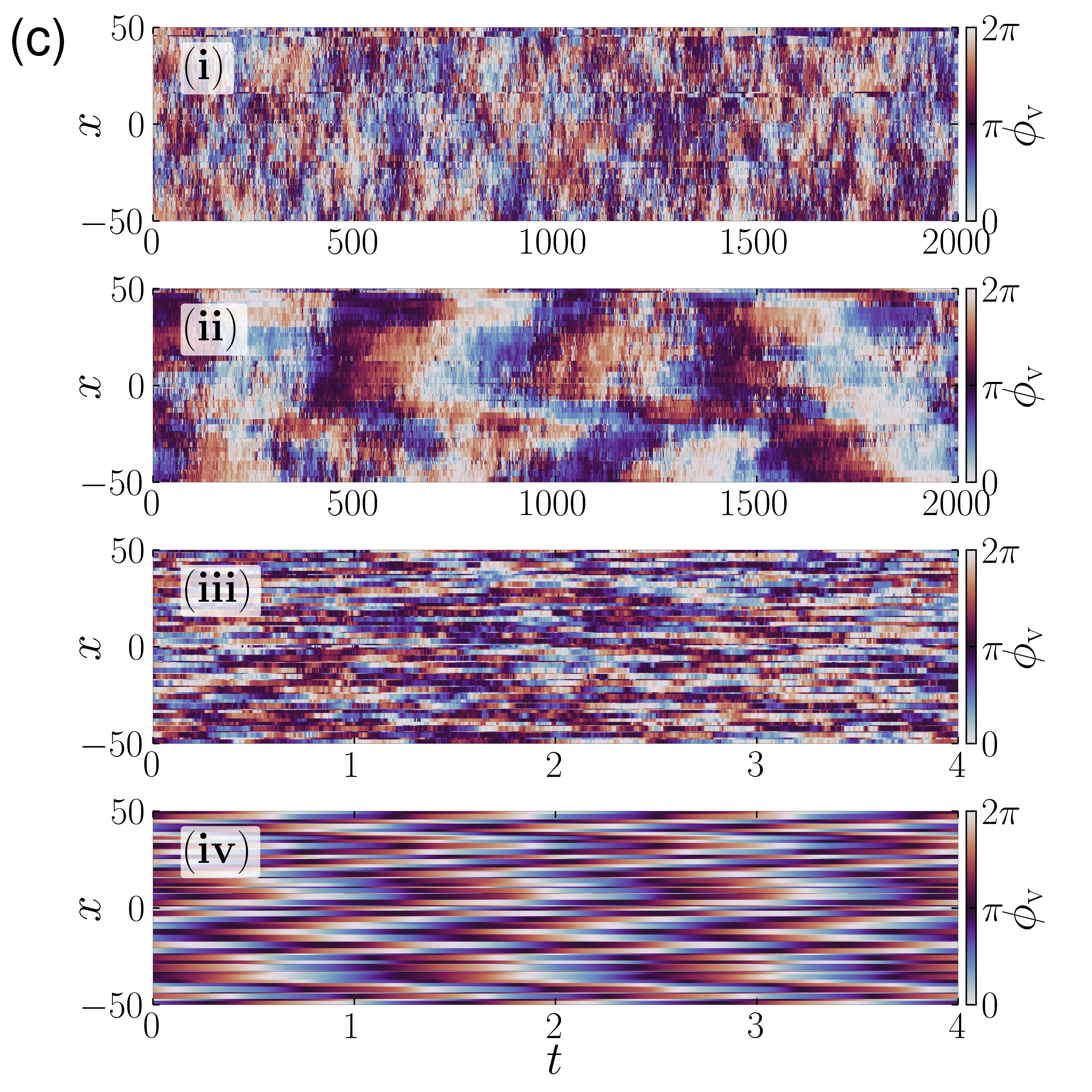}
\caption{Dynamical regimes and examples of states identified in active solids with noisy chiral dynamics.
(a) Phases on the $D_r-\W$ plane: Mesoscopic Range Order (MRO), Chiral Mesoscopic Range Order (CMRO), Dynamic Disorder (DD), and Chiral Disorder (CD). The thick red solid line corresponds to $D_r=\W$, separating the chiral ($D_r<\W$) and achiral ($D_r\geq\W$) regimes. 
The thin blue solid line represents $\xi_{\text{T}}=l_0$, where $l_0$ is the equilibrium distance between two neighboring particles.
The green dashed line indicates the maximum of total energy, showing re-entrant behavior from low energy to high energy and back to low energy along $D_r$ at constant high chirality. The curve with green open diamonds corresponds to $D_r^{*}=\W-\l_{max}$, where $\l_{max}$ is the maximum eigenvalue. 
The magenta dashed line represents the minimum of mean-squared velocity~(MSV), illustrating re-entrant behavior from high MSV to low MSV and then back to high MSV along $D_r$ at constant high chirality.
The curve with black open circles corresponds to the maximum of steady-state mean-squared displacement for a single harmonically trapped particle, denoted by $D_r^{*}=\W-\mu k$~(see Appendix-\ref{app:single_chiral_ht_MSD}).
(b) Snapshots in four distinct regimes: MRO, CMRO, DD, and CD. Complementing these, (c) showcases their respective kymographs, which depict space-time plots of the velocity angles, $\phi_\vv = \tan^{-1}(v_y/v_x)$, as obtained from simulations. These simulations correspond to the cross symbols marked in (a).
}
\label{fig1}
\end{figure*}     

\section{Model}
\label{sec_II}
We consider a system of $N$ densely packed soft chiral self-propelled disks following overdamped dynamics in a two dimensional periodic box of size $L \times L$.
Disregarding passive translational diffusion, the dynamics of the position $\rv_i\equiv(x_i,y_i)$ and heading direction $\nv_i\equiv(\cos(\th_i),\sin(\th_i))$ of the $i$-th disk will be given by
\bea
\dot{\rv}_i &=& v_0 \nv_i + \mu \Fv_i~,
\label{eom1}\\
\dot{\nv}_i &=& \left[ \W + \sqrt{2D_r}~\eta_i(t) \right] \nvperp_i~.\label{eom2}
\eea
Here, $v_0$ is the self-propulsion speed, $\mu$ is the mobility (inverse damping coefficient), $\W$ is the chiral angular speed of the disks, $D_r$ is their rotational diffusion coefficient, and $\nvperp_i$ is a unit vector perpendicular to $\nv_i$.
Noise is introduced through the random variable $\eta_i(t)$, following Gaussian white noise with $\la \eta_i (t)\ra=0$ and 
$\la\eta_i(t)\eta_j(t^{\prime})\ra=\d_{ij}\d(t-t^{\prime})$. 
The sum of all contact forces over disk $i$ is  
$\Fv_i = \sum_{j \in S_i} \fv_{ij}$,
where $S_i$ is the set of indexes of all disks that overlap disk $i$.
These forces are modeled as linear repulsion, with
$\fv_{ij} = k (|\rv_{ij}|-l_0)\rv_{ij}/|\rv_{ij}|$ if $|\rv_{ij}|\leq l_0$
and $\fv_{ij} = 0$ otherwise, were $\rv_{ij}=\rv_{j}-\rv_{i}$ and $l_0 = 2 r_0$ is the equilibrium center-to-center distance between two neighboring disks of radius $r_0$.
We note that, in real-world scenarios, $\W$ and $D_r$ will not be the same for all disks.
In Section~\ref{sec_VII}, we thus conduct a comprehensive investigation into binary and complex mixtures of disks with different $D_r$ and $\W$ values, substituting $D_r$ by $D_r^{i}$ and $\W$ by $\W^{i}$ in Eq.~(\ref{eom2}).
%

We note that the orientation dynamics in Eq.~(\ref{eom2}) are decoupled from the position dynamics in Eq.~(\ref{eom1}), and result from the interplay between deterministic chirality and angular diffusion.
The deterministic angular speed $\Omega$ sets a rotational timescale 
$\tauomega=\Omega^{-1}$; the diffusion constant $D_r$ sets a persistence timescale $\taur = D_r^{-1}$.
We will show below that the interplay between rotational, persistence, and elastic timescales can generate different collective states.

\section{State Space Overview}
\label{sec_III}

We begin by characterizing the different regimes that can be reached by the model introduced above.
Figure \ref{fig1}(a) presents a diagram of the resulting phases as a function of the chiral angular speed $\W$ and the angular diffusion coefficient $D_r$, with the boundaries computed analytically as we will detail in Section \ref{sec_IV}.
Broadly speaking, the system develops mesoscopic range order for low enough $\W$ and $D_r$ values (below the blue line), where patches of disks with strong velocity correlations spontaneously appear at different scales.
For high $\W / D_r$ ratios (above the red line) the velocity directions displayed by these patches rotate with a clearly defined chirality, determined by $\W$, defining the Chiral Mesoscopic Range Order (CMRO) regime. For low $\W / D_r$ ratios, no clear chirality is observed and we define the Mesoscopic Range Order (MRO) regime.
In the high $\W$ and high $D_r$ regimes (beyond the blue line), we find instead no extended regions of high velocity correlation. The individual particle motion is dominated by deterministic chiral rotation for high $\W / D_r$ ratios (above the red line), in the Chiral Disorder (CD) regime, and by stochastic rotational diffusion for low $\W / D_r$ ratios (below the red line), in the Dynamic Disorder (DD) regime.

Figures~\ref{fig1}(b) and \ref{fig1}(c) present snapshots of the velocity vectors and kymographs, respectively, describing the spatiotemporal dynamics of the velocity angles, for simulations in each one of the four regimes. 
Here the sub-panels correspond to:
(i) the MRO regime for $D_r=10^{-2}$, $\W=10^{-3}$, see Supplemental Material, Movie 1~\cite{Supply2024};
(ii) the CMRO regime for $D_r=10^{-3}$, $\W=10^{-2}$, see Movie 2~\cite{Supply2024}; 
(iii) the DD regime for $D_r=5$, $\W=10^{-2}$, see Movie 3~\cite{Supply2024}; and 
(iv) the CD regime for $D_r=10^{-2}$, $\W=5$, see Movie 4~\cite{Supply2024}.

All simulations were carried out for $N=3183$ disks of radius $r_0 = 1$ in a periodic square box of side $L=100$, which results in a packing fraction of $\phi = N \pi r_0^2 / L^2 \approx 1$, and for the following simulation parameters (unless otherwise stated): mobility $\mu=1$, elastic repulsive strength $k=1$, and active speed $v_0=0.01$. 
Spatially, they form a crystalline triangular packings without defects, well in the solid phase, without rearrangements for the duration of the simulation.
In the snapshots, each disk is represented by a small arrow starting at $\rv_i$, pointing towards $\dot{\rv}_i$, with length proportional to $\lVert \dot{\rv}_i \rVert$, and colored by angle. 
In the kymographs, we use colors to display the angle of the velocity $\dot{\rv}_i$ of all disks located within a narrow slit, with $-r_0 \leq y \leq +r_0$, as a function of their $x$ position and time.

The snapshots in sub-panels (i) and (ii) of Fig.~\ref{fig1}(b) clearly show the emergence of mesoscopic-range order, while the kymographs in sub-panels (i) and (ii) of Fig.~\ref{fig1}(c) show that their temporal dynamics is distinct, with only sub-panel (ii) showing periodic dynamics that result from a close to deterministic local rotation of the $\hat{n}_i$ vector. 
Correspondingly, the snapshots in sub-panels (iii) and (iv) of Fig.~\ref{fig1}(b) show disordered states, while the kymographs in sub-panels (iii) and (iv) of Fig.~\ref{fig1}(c) show that the dynamics in the DD regime is random in time while the CD regime dynamics is quasiperiodic.
Note that the periodicity of the angular dynamics in sub-panels (ii) and (iv) of Fig.~\ref{fig1}(c) matches the expected full rotation period $T= 2\pi/\W$, with $T=200 \pi  \simeq 628.32$ for (ii) and $T=2 \pi / 5 \simeq 1.26$ for (iv). 

In addition to the four regimes described above, the diagram in Fig.~\ref{fig1}(a) also contains dashed green and magenta lines, as well as lines of open black circles and green diamonds. These trace four different analytical approximations for the location in the diagram of the `hammering state' identified in~\cite{Debets2023}, where the elastic energy contained by the system is maximal and its kinetic energy is minimal (see Supplemental Material, Movie 5~\cite{Supply2024} for a simulation with $D_r = \W = 10$).

We will deduce analytically below the dynamics and boundaries of the different regimes described above.

\begin{figure*}[!ht]
\includegraphics[width=8.50cm]{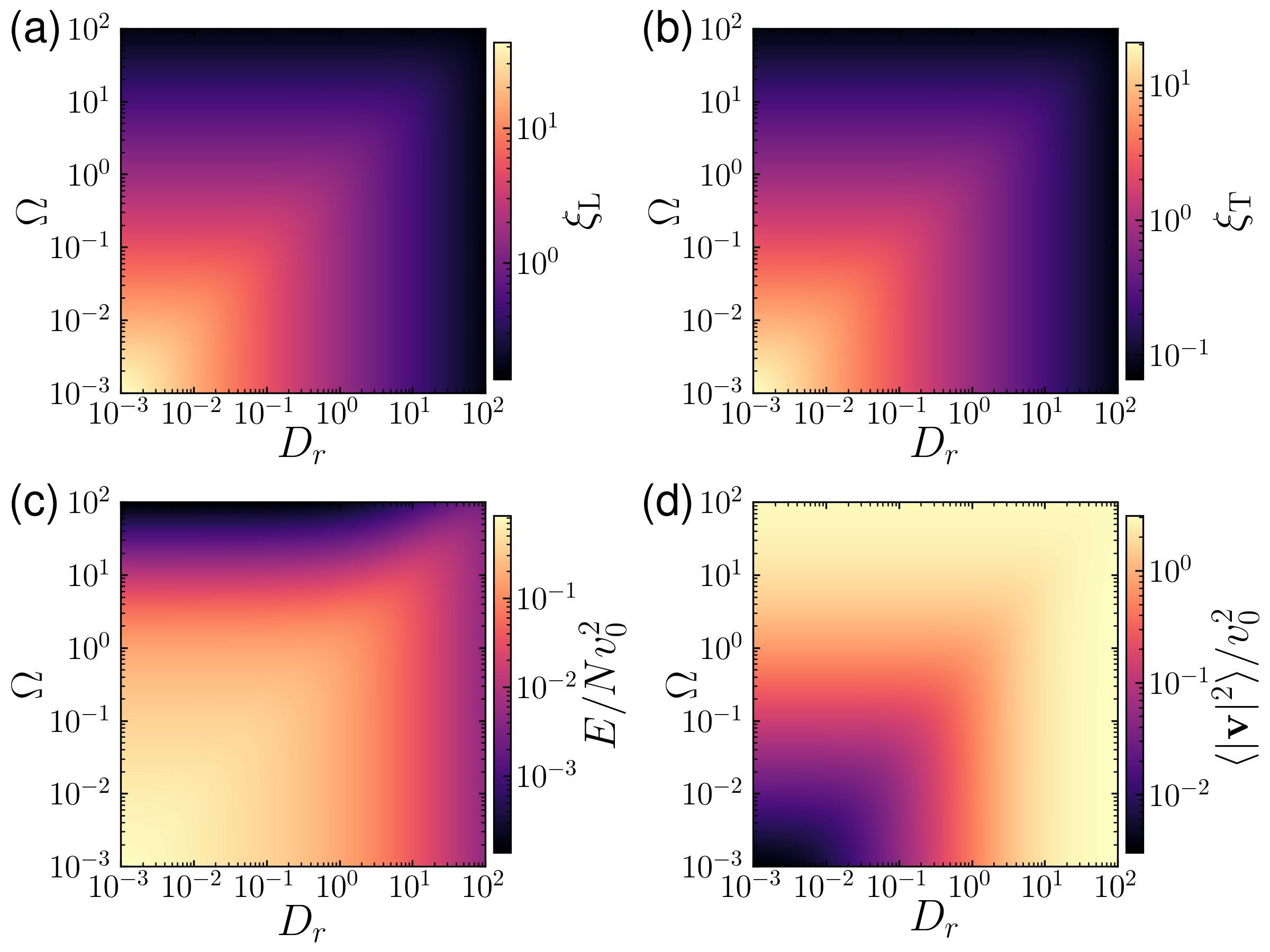}
\includegraphics[width=8.50cm]{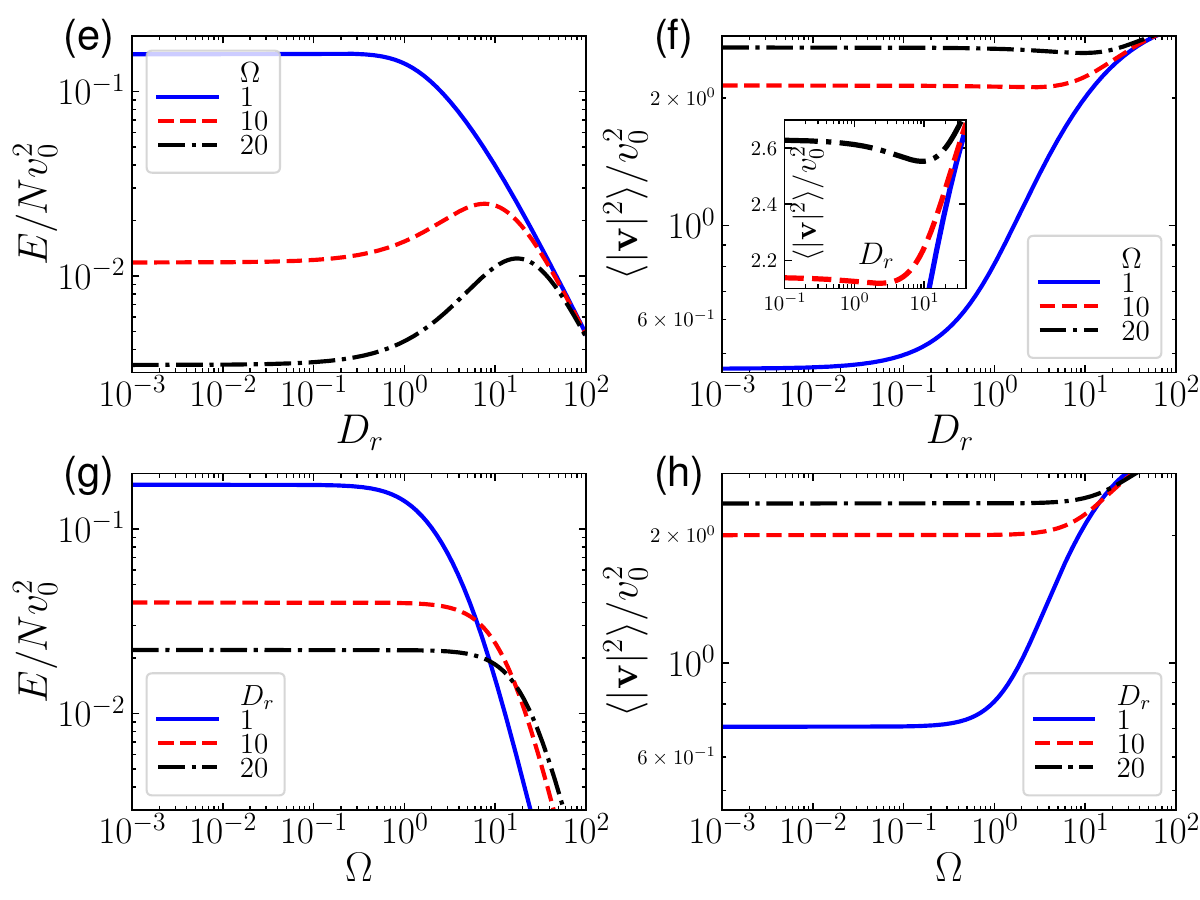}
\caption{Analytic Behavior of Characteristic Length Scales, Energy, and Mean-Squared Velocity. (a), (b) Color maps of the longitudinal~$\xi_{\text{L}}$ and transverse~$\xi_{\text{T}}$ characteristic length scales on the $D_r-\W$ plane, respectively, defined using the continuum elastic formulation Eq.~(\ref{eq:characteristic_length_scales}).
(c) Color map of the energy $E/Nv_0^2 = \sum_{\nu} E_{\nu}/Nv_0^2$ on the $D_r-\W$ plane, as derived from the normal mode formulation Eq.~(\ref{eq:avg_energy_per_mode}).
(d) Color map of the mean-squared velocity $\langle |\mathbf{v}|^2 \rangle/v_0^2$ on the $D_r-\Omega$ plane resulting from the continuum elastic formulation Eq.~(\ref{eq:MSV}). (e) $E/Nv_0^2$ as a function of $D_r$ for $\W=1,10,20$, and (g) $E/Nv_0^2$ as a function of $\W$ for $D_r=1,10,20$. (f) $\langle |\mathbf{v}|^2 \rangle/v_0^2$ as a function of $D_r$ for $\W=1,10,20$, and (h) $\langle |\mathbf{v}|^2 \rangle/v_0^2$ as a function of $\W$ for $D_r=1,10,20$. Inset in (f) depicts the minimum of $\langle |\mathbf{v}|^2 \rangle/v_0^2$ at intermediate $D_r$ values for high $\W$ values.
}
\label{fig2}
\end{figure*}     

\section{Analytical Results}
\label{sec_IV}

In this section, we present the analytical formulations used to describe the system.
We begin by computing the orientation dynamics of the heading direction in Subsection A, since they are not coupled to the positions.
We then formulate a linear response theory, adopting the method in Henkes et al.~\cite{Henkes2020} to describe the linear response in terms of normal modes in Subsection B, to then calculate the energy per mode and spatial velocity correlations.
We further simplify the analytical description by implementing a continuum elasticity framework in Subsection C, to compute mean-squared velocity and velocity autocorrelation functions.

\subsection{Orientation Dynamics}
\label{subsec_orientation_autocorr}
Given that the orientation evolves independently in Eq.~(\ref{eom2}), the probability distribution $P(\nv,t)$ for the heading direction $\nv$ as a function of time will follow the Fokker-Planck equation
\bea
\p_{t} P(\nv,t) &=& D_r \nabla_{\nv}^{2} P -\W  \nv^{\perp} \cdot \nabla_{\nv} P~,
\label{F-P-orientation}
\eea
where $\nabla_{\nv}$ is the Laplacian in orientation space. 
Using a Laplace transformation approach described in detail in Appendix-\ref{app:orientation_autocorr}, we can compute an exact expression for the heading orientation autocorrelation, which is given by
\bea
\la\nv(t)\cdot\nv(0)\ra &=& \text{e}^{-D_r t} \cos(\Omega~t).
\label{eq:ncorr}
\eea

Here, the decay rate of the exponential term is given by $\tau_r = D_r^{-1}$ and the period of chiral rotation, by $\tauomega = \W^{-1}$. 
We thus define $D_r=\W$ as the critical line between a regime dominated by the angular noise and a regime dominated by the deterministic chirality, which we highlighted as a solid red line in Fig.~\ref{fig1}(a). 
Note that this boundary is formally analogous in Eq.~(\ref{eq:ncorr}) to the limit between damped and overdamped oscillations, where the high angular diffusion case corresponds to the overdamped regime, as the mean temporal heading correlations display no oscillatory component.

\subsection{Normal Mode Formulation}
\label{subsec_normal_mode_formulation}

In order to express the dynamics in terms of the normal modes of vibration of the passive system, we first define as $\rv_i^0$ the equilibrium position of disk $i$ for $v_0=0$, which corresponds to a minimum of the elastic energy.
Using Eq.~(\ref{eom1}), we then find that the dynamics of small displacements $\d\rv_i = \rv_i - \rv_i^0$ around these equilibrium positions are described by
\bea
\d\dot{\rv}_i &=& v_0 \nv_i - \sum_{j} \mathbb{K}_{ij} \cdot \d\rv_{j} ,
\label{eq:linearize_eom0}
\eea
where each $\mathbb{K}_{ij}$ corresponds to a $2\times 2$ block of the $2N\times 2N$ dynamical matrix. 
We are interested in expressing the dynamics over the normal elastic modes of the system, i.e., over the eigenvectors of the dynamical matrix.
Each of these $2N$ normal modes corresponds to a $2N$-dimensional eigenvector that can be written as a list of $N$ two-dimensional vectors, given by~$({\bf\xi}_{1}^\nu,...,{\bf\xi}_{N}^\nu)$, where $\nu=1,...,2N$ labels the eigenvector mode associated to the eigenvalue $\l_\nu$.

We can formally write the displacements in terms of the eigenmodes described above as
$\d \rv_i = \sum_{\nu=1}^{2N} a_{\nu} \xiv_{i}^{\nu}$.
Projecting Eq.~(\ref{eq:linearize_eom0}) onto the normal modes, we then obtain the following uncoupled equations for the dynamics of the normal mode amplitudes:
\bea
\dot{a}_{\nu} &=& \eta_{\nu} - \l_{\nu} a_{\nu},
\label{eq:normal_modes}
\eea
where $\eta_{\nu}$ is the projection of the self-propulsion force onto the normal mode $\nu$, given by
\bea
\eta_{\nu} &=& v_0 \sum_{i=1}^{N} \nv_i \cdot \xiv_i^{\nu}.
\label{eq:noise_mode}
\eea
We note that $\eta_{\nu}$ is the sum of $N$ statistically independent contributions with bounded moments, each one resulting from the correlated noise dynamics in time that is followed by its corresponding $\nv_i$.
The Central Limit Theorem then implies that $\eta_{\nu}$ must follow a Gaussian distribution, here with $\la \eta_{\nu} (t)\ra = 0$.
Additionally, since the eigenvectors form an orthonormal basis where 
$\sum_{i=1}^{N}\xiv_i^{\nu}\cdot\xiv_i^{\nu^{\prime}}=\d_{\nu,\nu^{\prime}}$, 
the corresponding two-time correlation function will be given by
$\la \eta_{\nu} (t) \eta_{\nu^{\prime}} (t^{\prime})\ra = (v_0^2/2) \la\nv(t)\cdot\nv(t^{\prime})\ra \d_{\nu,\nu^{\prime}}$.
Replacing the heading autocorrelation expression in Eq.~(\ref{eq:ncorr}), we finally obtain
$\la \eta_{\nu} (t) \eta_{\nu} (0)\ra = (v_0^2/2) \text{e}^{-D_r t} \cos{(\W~t)}$, which implies that the statistical properties of the noise $\eta_{\nu}$ are the same for any mode $\nu$.

We now calculate the mean potential energy stored in each mode (see Supplemental Material, Section II for details \cite{Supply2024}).
By solving Eq.~(\ref{eq:normal_modes}), we first find
\bea
a_{\nu}(t) &=& a_{\nu}(0) \text{e}^{-\l_{\nu}t} + 
 \int_{0}^{t} dt^{\prime} \eta_{\nu} (t^{\prime}) \text{e}^{-\l_{\nu}(t-t^{\prime})}~.
 \label{eq:anu_time_expression}
\eea
From here, we can obtain the steady state mean squared value of $a_{\nu}(t)$ by computing
$\Lim \la a_{\nu}^2 (t) \ra$ to obtain
\bea
\la a_{\nu}^2\ra &=& \frac{v_0^2 (D_r  +\l_{\nu})}{2\l_{\nu} \left[(D_r  +\l_{\nu})^2 + \W^2 \right]}~. 
\label{eq:anu2_average}
\eea
The mean energy per mode is given by $E_{\nu} = \l_{\nu}\la a_{\nu}^2\ra/2$ and can thus be expressed as
\bea
E_{\nu} &=& \frac{v_0^2 (D_r  +\l_{\nu})}{4 \left[(D_r  +\l_{\nu})^2 + \W^2 \right]}~. 
\label{eq:avg_energy_per_mode}
\eea
We note in this equation that there is a critical curve in the $D_r-\W$ plane that maximizes the mean potential energy injected into the system by the combined activity of all modes (as shown in Figure S2 of the Supplemental Material \cite{Supply2024}).
We can obtain an approximate expression for this curve by finding the conditions that maximize the energy of the stiffest mode only (i.e., the mode least excited by the activity), which we identified as the main responsible for the maximum in the total potential energy.
Since the stiffest mode corresponds to the largest eigenvalue $\l_{\nu}=\l_{max}$, its energy will be $E^*=E_{\nu}|_{\l_{\nu}=\l_{max}}$ and its maximum can be computed using $\p E^*/\p D_r = 0$.
We thus find that the potential energy injected by activity is approximately maximized for
$D_r^{*}=\W-\l_{max}$, corresponding to the dominant mode, i.e. the maximum eigenvalue $\lambda_{max}=5.93\pm0.01$.
This curve is displayed as the green open diamonds in Fig.~\ref{fig1}(a).

In Fig.~\ref{fig2}(c), we visualize a color map of the energy $E/Nv_0^2 = \sum_{\nu} E_{\nu}/Nv_0^2$ on the $D_r- \W$ plane that clearly shows an increase of elastic energy in the low $D_r$ and low $\W$ regimes.
Figure~\ref{fig2}(e) presents $E/Nv_0^2$ as function of $D_r$ for three different $\W=1,10,20$ values, showing the presence of a maximum at intermediate $D_r$ noise strengths, for high chirality ($\W=10,20$). In this regime, we thus find that the elastic energy can grow despite an increase in noise strength.
In Fig.~\ref{fig2}(g), we plot $E/Nv_0^2$ as function of $\W$ for three different $D_r=1,10,20$ values, showing a monotonic decrease of the elastic energy. We display the maximum of $E/Nv_0^2$ in the $D_r-\W$ plane as the green dashed line in Fig.~\ref{fig1}(a), which matches the previously computed $D_r^{*}=\W-\l_{max}$ curve.

Equation (\ref{eq:avg_energy_per_mode}) also provides us with expressions for the 
low and high limits of angular noise or chirality.
In the high noise case, $D_r/\lambda_\nu \gg 1$ and the mean energy per mode reduces to $E_{\nu} \approx v_0^2 D_r/4(D_r^2 +\W^2)$, which gives rise to two limits:
(\text{i}) a low chirality limit~$\W\to 0$, where $E_{\nu} \to v_0^2/4D_r$, and
(\text{ii}) a high chirality limit~$\W\to\infty$, where $E_{\nu} \to 0$.
%
In the low noise case $D_r\to 0$, we find $E_{\nu}=v_0^2/4[\l_{\nu}^2 +\W^2]$, which also gives rise to two limits: 
(\text{i}) a low chirality limit with $E_{\nu} \to v_0^2/4\l_{\nu}^2$, where the lowest modes with $\l_{\nu} \ll 1$ are enhanced,
and (\text{ii}) a high chirality limit with $E_{\nu} \to v_0^2/4\W^2$.
%
On the other hand, for any noise value, in the $\W\to 0$ limit, we recover from Eq.~(\ref{eq:avg_energy_per_mode})  the same expressions previously obtained in~\cite{Bi2016, Henkes2020} for standard (non-chiral or achiral) active particles, as expected.

Finally, in order to identify the emergence of mesoscopic order, we are interested in finding the mean velocity spectrum (see Supplemental Material, Section II A for details \cite{Supply2024}).
We begin by expressing the velocity in Fourier space, computing its discrete Fourier transform 
$\vv(\qv) = \sum_{j=1}^{N} \text{e}^{i\qv\cdot\rv_{j}^{0}} \delta\dot{\rv}_{j}/N$
in terms of the $\rv_{i}^{0}$ equilibrium reference positions of the disks.  
Expanding $\delta\dot{\rv}_{j}$ in the normal mode basis, we find
\begin{eqnarray}
\la|\vv(\qv)^2|\ra &=& \la \vv(\qv)\cdot\vv^{*}(\qv)\ra 
= \sum_{\nu,\nu^{\prime}} \la \dot{a}_{\nu}\dot{a}_{\nu^{\prime}}\ra \xiv_{\nu} (\qv) \cdot \xiv^{*}_{\nu^{\prime}} (\qv) \nonumber \\
\label{eq:velSpec1}
&=& \sum_{\nu,\nu^{\prime}} \la \dot{a}_{\nu}^2\ra |\xiv_{\nu} (\qv)|^2 \d_{\nu\nu^{\prime}},
\end{eqnarray}
where we defined
$\xiv_{\nu} (\qv) = \sum_{j=1}^{N} \text{e}^{i\qv\cdot\rv_{j}^{0}} \xiv_{j}^{\nu}/N$
as the discrete Fourier transform of the eigenvectors.
Using Eq.~(\ref{eq:normal_modes}), we then replace
$\la \dot{a}_{\nu}^2\ra =
\l_{\nu}^2\la a_{\nu}^2\ra - 2 \l_{\nu} \la a_{\nu}\eta_{\nu}\ra + \la \eta_{\nu}^2\ra$
into Eq.~(\ref{eq:velSpec1}).
Here, the $\la a_{\nu}^2\ra$ term is known from Eq.~(\ref{eq:anu2_average}), 
the equal-time correlation $\la\eta_{\nu}^2\ra=v_0^2/2$ can be computed from 
Eq.~(\ref{eq:noise_mode}), 
and the expression for 
$\la a_{\nu}\eta_{\nu}\ra=v_0^2(D_r+\l_{\nu})/2[(D_r+\l_{\nu})^2+\W^2]$ 
in the steady-state ($t\to\infty$) 
can be obtained from Eq.~(\ref{eq:anu_time_expression}).
This leads to the following explicit expression for the velocity correlation function:
\begin{equation}
\la|\vv(\qv)^2|\ra \! = \!\f{v_0^2}{2} \sum_{\nu} \left[  1 \!-\! \frac{\l_{\nu}(D_r + \l_{\nu})}{(D_r \!+\! \l_{\nu})^2 \!+\! \W^2} \right] \!|\xiv_{\nu}(\qv)|^2.
\label{eq:spatial_vel_corr_mode}
\end{equation}
This equation allows us examine different limits. 
For $\Omega=0$, it simplifies to the velocity correlation function 
$\la|\vv(\qv)^2|\ra = (v_0^2/2) \sum_{\nu} [D_r/(D_r+\l_{\nu})] |\xiv_{\nu}(\qv)|^2$, 
previously obtained for achiral active particles in \cite{Henkes2020}. 
For $D_r=0$, it simplifies to the velocity correlation function $\la|\vv(\qv)^2|\ra = (v_0^2/2) \sum_{\nu} [\W^2/(\W^2+\l_{\nu}^2)] |\xiv_{\nu}(\qv)|^2$ for disordered deterministic rotators.

In most experimental contexts, the extraction of the normal modes or their eigenvalues is unfeasible, except in specific scenarios like colloidal particle experiments~\cite{Chen2010, Henkes2012,melio2024soft}.
Current methods are often restricted to measuring in thermal equilibrium conditions and necessitate extensive data gathering. In the next subsection, we will therefore extend our findings to the framework of continuum elasticity theory, which only requires knowing the elastic constants of the material and is thus much easier to compute for real-world systems.

\begin{figure*}[!ht]
\includegraphics[width=7cm]{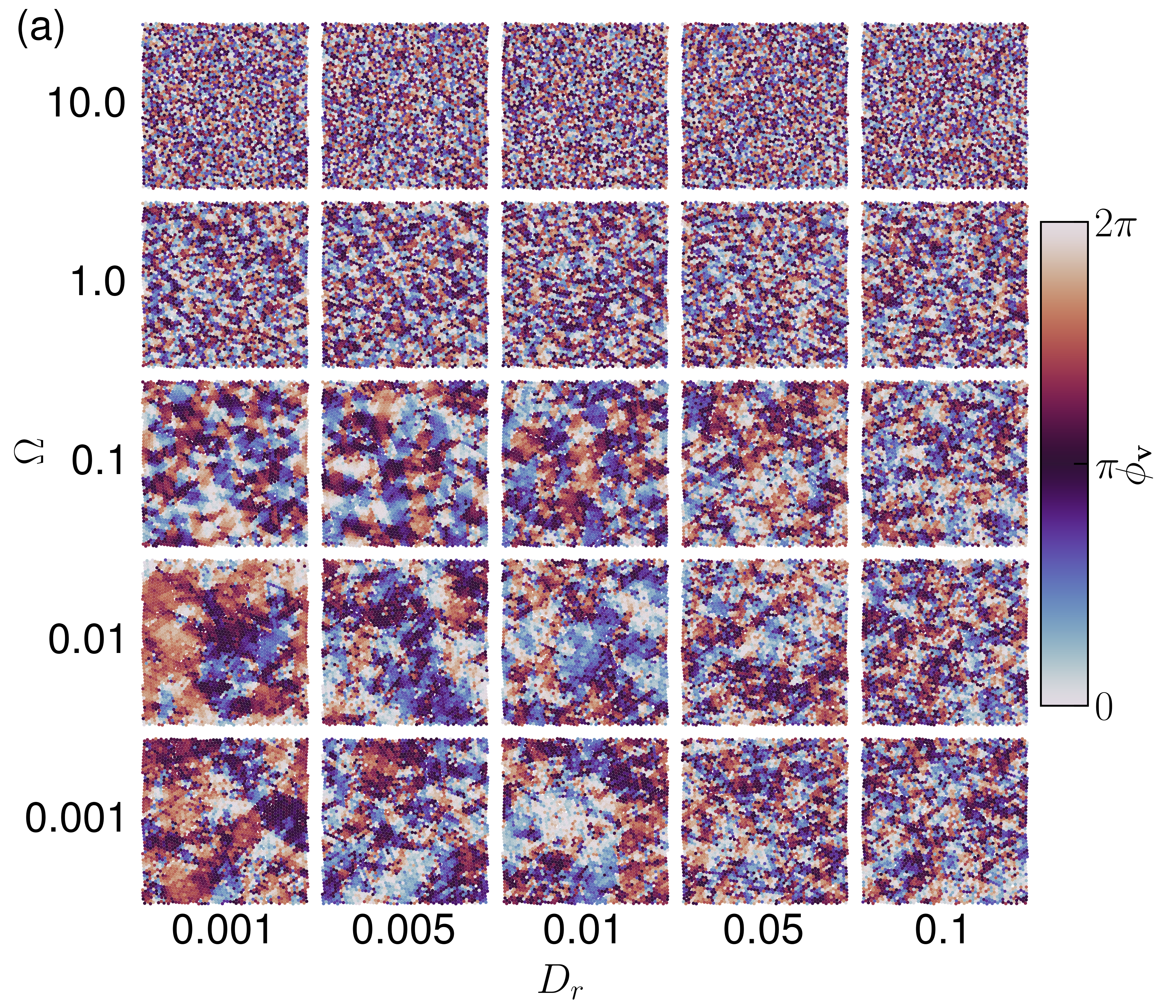}
\includegraphics[width=10cm]{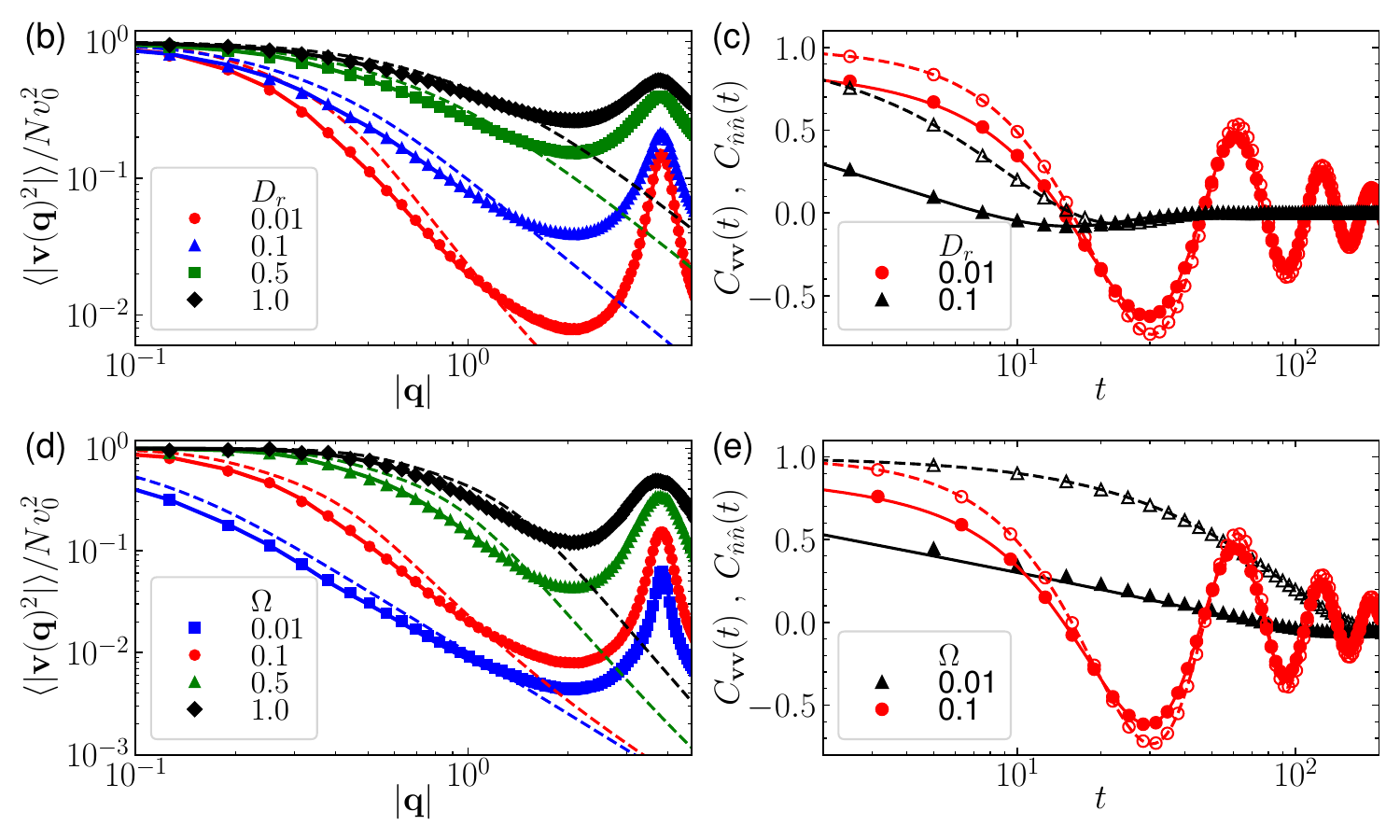}
\caption{Chiral Mesoscopic Range Order at Low Chirality and Low Noise. (a) Snapshots in the $\Omega-D_r$ parameter space illustrate the emergent correlations through the distribution of velocity angles $\phi_\vv = \tan^{-1}(v_y/v_x)$.
(b),(d) Spatial velocity correlation function in Fourier space $\la|\vv(\mathbf{q})^2|\ra/N v_0^2$ as a function of $\qv$, from simulation (symbols), while solid lines are results from the normal mode formulation Eq.~(\ref{eq:spatial_vel_corr_mode}) and dashed lines correspond to the continuum elastic formulation Eq.~(\ref{eq:spatial_vel_corr_continuum}). (b) $D_r=0.01$~(red circles),~$0.1$~(blue triangles),~$0.5$~(green squares),~$1.0$~(black diamonds) with $\W=0.1$. (d) $\W=0.01$~(blue squares),~$0.1$~(red circles),~$0.5$~(green triangles),~$1.0$~(black diamonds) with $D_r=0.01$.
(c),(e) Normalized velocity autocorrelation functions $C_{\vv\vv}(t)=\la\vv(t)\cdot\vv(0)\ra/\la\vv(0)^2\ra$ (solid lines and filled symbols) and orientation autocorrelation functions $C_{\nv\nv}(t)=\la\nv(t)\cdot\nv(0)\ra$ (dashed lines and open symbols) as a function of time $t$. Symbols are simulation results, while solid lines correspond to the $C_{\vv\vv}(t)$ results from the continuum elastic formulation Eq.~(\ref{eq:velocity_autocorrelation_continuum}), while dashed lines are the plot of Eq.~(\ref{eq:ncorr}). (c) $D_r=0.01$~(red circles),~$0.1$~(black triangles) with $\W=0.1$. (e) $\W=0.01$~(black triangles),~$0.1$~(red circles) with $D_r=0.01$.  
}
\label{fig3}
\end{figure*}

\subsection{Continuum Elastic Formulation}
\label{subsec_continuum_elastic_formulation}
To derive the continuum formulation, we begin by writing the  equation of motion for the displacement vector field $\uv(\rv)=\rv^{\prime}(\rv)-\rv$, which describes the deformed state $\rv^{\prime}(\rv)$ with respect to the equilibrium reference state $\rv$.
As detailed in the Supplemental Material~\cite{Supply2024}, in the presence of active forces this equation is given by
\bea
\dot{\uv} &=& \nabla \cdot \mathcal{\sigma} + \fv_{\text{act}}.
\label{eq_continuum_eom}
\eea
Here, $\s$ is the passive stress tensor, with components $\sigma_{\a\b} = B \d_{\a\b} u_{\g\g} +2 G (u_{\a\b} - \frac{1}{2} \d_{\a\b} u_{\g\g})$,
and activity is introduced through self-propulsion forces defined by
$\fv_{\text{act}}(\rv,t)=v_0 \nv(\rv,t)$.
In this expression for the stress tensor, $B$ and $G$ correspond  respectively to the bulk and shear moduli of the isotropic solid,
the strain tensor components 
$u_{\a\b}=\frac{1}{2}\left[\partial_{\a}u_{\b}+\partial_{\b}u_{\a}\right]$ 
are written in terms of spatial derivatives of the displacement vectors $\uv(\rv)$ with respect to $\a,\b\in \{x,y\}$, and the summation over repeated indexes is assumed.
We can see explicitly in Eq.~(\ref{eq_continuum_eom}) that this active solid is distinct from odd active matter, which only considers internal active stresses that can be written in terms of effective moduli \cite{fruchart2023odd}.

To proceed with the computations, we define the direct and inverse spatiotemporal Fourier transforms as
\begin{eqnarray}
\uv(\rv,t) &=& \frac{1}{(2\pi)^3} \int d^2\qv \int d\omega ~\tilde{\uv} (\qv,\omega) \text{e}^{-i(\qv\cdot\rv+\omega t)} ,\nonumber \\
\tilde{\uv}(\qv,\omega) &=&  \int d^2\rv \int dt ~\uv(\rv,t) \text{e}^{i(\qv\cdot\rv+\omega t)}, \nonumber
\end{eqnarray}
and write the continuum equation of motion~(\ref{eq_continuum_eom}) in Fourier space as 
\bea
-i \omega \tilde{\uv} (\qv,\omega) &=& \tilde{\fv}_{\text{act}} (\qv,\omega) -\mathbb{D}(\qv) \tilde{\uv}(\qv, \omega).
\label{eq:Fourier_space_eom}
\eea
Here, $\mathbb{D}(\qv)$ is a $2\times 2$ dynamic matrix in Fourier space, given by
$$\mathbb{D}(\qv) = \begin{bmatrix}
B q_x^2 + G q^2 & B q_x q_y \\
B q_x q_y & B q_y^2 + G q^2 \\
\end{bmatrix},$$ 
where $q^2=q_x^2+q_y^2$ (see Supplemental Material~\cite{Supply2024} for a detailed derivation), and we defined the active force $\fv_{\text{act}}(\rv,t)$ in Fourier space as
\bea
\tilde{\fv}_{\text{act}} (\qv,\omega)  &=& v_0 \int d^2\rv \int_{-\infty}^{\infty} dt ~\nv(\rv,t) ~\text{e}^{i(\qv\cdot\rv+\omega t)}~.
\label{eq:Fourier_transform_active_force}
\eea 

We are interested in computing the velocity correlation functions. To do this, we begin by writing the orientational correlation in the continuum limit, replacing $\nv_i(t)$ by a continuous field $\nv(\rv, t)$, with $\la\nv_i(t)\cdot \nv_j(t^{\prime})\ra = \d_{i,j} \la\nv(t)\cdot \nv(t^{\prime})\ra$.
We then substitute the Kronecker delta $\d_{i,j}$ by its Dirac counterpart, using $\d_{i,j} \to a^2 \d(\rv-\rv^{\prime})$, where $a$ is the smallest characteristic length scale of the system, 
to obtain $\la\nv(\rv, t)\cdot \nv(\rv^{\prime}, t^{\prime})\ra = a^2 \d(\rv-\rv^{\prime}) \la\nv(t)\cdot \nv(t^{\prime})\ra$.
From Eq.~(\ref{eq:Fourier_transform_active_force}), it is then clear that $\la \tilde{\fv}_{\text{act}} (\qv,\omega)   \ra = 0$, and the second order correlation function $C_{\tilde{F}} = \la \tilde{\fv}_{\text{act}} (\qv,\omega) \cdot \tilde{\fv}_{\text{act}} (\qv^{\prime},\omega^{\prime}) \ra$ is simply given by 
\bea
C_{\tilde{F}} &=&   \frac{2(2\pi)^3 a^2 v_0^2 D_r}{(\omega-\Omega)^2+D_r^2} \d(\qv+\qv^{\prime}) \d(\omega+\omega^{\prime})~.
\label{eq:force_corr_continuum}
\eea
If we now consider a finite system (a square of side $L$, for simplicity), the wave vector becomes discretized. 
We can thus replace the Dirac delta by the Kronecker delta, $\d(\qv+\qv^{\prime}) \to \frac{1}{(\Delta q)^2} \d_{\qv^{\prime},-\qv}$,
with $\Delta q \equiv 2\pi/L$. 
This also leads us to define the spatially discrete  Fourier transform $\fv_{\text{act}} (\qv,\omega)  = \tilde{\fv}_{\text{act}} (\qv,\omega) /a^2$ for discrete spatial wave vectors $\qv$ but continuous frequency $\omega$. 
The correlation function for this discrete Fourier transform, given by $C_{F}= \la \fv_{\text{act}} (\qv,\omega)\cdot  \fv_{\text{act}} (\qv^{\prime},\omega^{\prime}) \ra$, will be equal to
\bea
C_{F} &=& \frac{N\pi^2v_0^2 D_r}{\phi \left[(\omega-\Omega)^2+D_r^2\right]}\d(\omega+\omega^{\prime})~.
\label{eq:force_corr_discrete}
\eea
Finally, by decomposing Eq.~(\ref{eq:Fourier_transform_active_force}) into a its 
longitudinal and transverse components 
$\tilde{\uv}=\tilde{u}_{\text{L}}(\qv, \omega) \hat{\qv} + \tilde{u}_{\text{T}} (\qv, \omega) \hat{\qv}_{\perp}$, with respect to the wave vector $\qv$, 
we can use $\tilde{\vv} (\qv, \omega) = -\text{i}\omega \tilde{\uv} (\qv, \omega)$ to obtain the following expression for the mean squared velocity in Fourier space
\bea
\la|\vv(\qv)|^2\ra &=& \frac{N v_0^2}{2} \left[\frac{1+ \chi (\xi_{\text{L}}q)^2}{1+2 \chi (\xi_{\text{L}}q)^2 + (\xi_{\text{L}}q)^4} \right.\nonumber\\
&+&\left. \frac{1+ \chi (\xi_{\text{T}}q)^2}{1+2 \chi (\xi_{\text{T}}q)^2 + (\xi_{\text{T}}q)^4} \right]~.
\label{eq:spatial_vel_corr_continuum}
\eea
Here, we have respectively defined the longitudinal and transverse characteristic length scales as
\bea
\label{Eq:xiLxiT}
\xi_{\text{L}} =\sqrt{\frac{B+G}{\sqrt{D_r^2+\W^2}}}~,~\xi_{\text{T}} = \sqrt{\frac{G}{\sqrt{D_r^2+\W^2}}},
\label{eq:characteristic_length_scales}
\eea
and the control parameter as $\chi=D_r/\sqrt{D_r^2+\W^2}$.
%

%
%
Figures \ref{fig2}(a) and \ref{fig2}(b) display how the longitudinal and transverse characteristic length scales described by Eqs.~(\ref{Eq:xiLxiT}) change across the different regimes on the $D_r-\W$ plane.
We observe that the largest characteristic length scales are found for small $D_r$ and $\W$ values.
Looking back at figure~\ref{fig1}(a), we now identify the thin blue line as the $D_r$ and $\W$ values for which the smallest characteristic length scale $\xi_{\text{T}}$ is equal to the typical equilibrium distance between particles $l_0$.
Below this line, the system develops mesoscopic scale correlations.

Next, we proceed to compute the mean-squared velocity $\la |\vv|^2\ra$ of our system in real space. 
We can directly write the mean-squared velocity of all particles, $\la |\vv|^2\ra=\la\sum_{i}|\vv_i|^2\ra/N$, in continuum form as $\la |\vv|^2\ra  = [ a^2/N(2\pi)^2] \int d^2\qv~\la|\vv(\qv)|^2\ra$, where $\la|\vv(\qv)|^2\ra$ was already evaluated in Eq.~(\ref{eq:spatial_vel_corr_continuum}). 
The upper limit of this integral is set by the inverse particle size, i.e., by the maximum wave number $q_\mathrm{max}=2\pi/a$ where $a$ is of the order of the particle size (this corresponds to $a=l_0$ for the simulation). Using $\int d^2q=2\pi\int q~dq$, we thus write 
\bea
\la |\vv|^2\ra = \frac{a^2}{2\pi N} \int dq~q~\la|\vv(\qv)|^2\ra.
\label{eq:MSV}
\eea
This expression can then be integrated numerically to compute the mean-squared velocity. In Fig.~\ref{fig2}(d), we visualize a color map of the mean-squared velocity $\la|\vv|^2\ra/v_0^2$ on the $D_r-\W$ plane, which shows that the mean-squared velocity is small for low $D_r$ and low $\W$ values, in regimes displaying mesoscopic coherent motion.

Figure~\ref{fig2}(f) presents $\la|\vv|^2\ra/v_0^2$ as function of $D_r$ for three different values of $\W=1,10,20$. Its inset shows the presence of a minimum in the $\la|\vv|^2\ra/v_0^2$ value at intermediate $D_r$ noise strengths, for high chirality ($\W=10,20$). This means that the kinetic energy is suppressed despite an increase in noise strength. Figure~\ref{fig2}(h) displays $\la|\vv|^2\ra/v_0^2$ as function of $\W$ for three different values of $D_r=1,10,20$.
Looking back at figure~\ref{fig1}(a), we plot the magenta dashed line as the minimum of the normalized mean-squared velocity $|\la\vv^2\ra|/v_0^2$ (as apparent in Fig.~\ref{fig2}(d)), where most of the self-propulsion becomes potential energy. 
This curve can be directly computed through a numerical integration of Eq.~(\ref{eq:MSV}).

Finally, we will investigate the collective temporal behavior of the system using the same continuum approach. Please see the Supplementary Material sections III B-C for details \cite{Supply2024}. We can directly calculate the velocity autocorrelation function defined by the integral over Fourier space as
\bea
\la \vv(t)\cdot \vv(t^{\prime})\ra &=& \f{1}{(2\pi)^4} \int d^2\qv \int d^2\qv^{\prime}~\la \tilde{\vv}(\qv,t)\cdot\tilde{\vv}(\qv^{\prime},t^{\prime}) \ra.\nonumber\\ 
\eea
Here, $\la \tilde{\vv}(\qv,t)\cdot\tilde{\vv}(\qv^{\prime},t^{\prime}) \ra$ can be written in terms of the Fourier integral of $\la \tilde{\vv}(\qv,\w)\cdot\tilde{\vv}(\qv^{\prime},\w^{\prime}) \ra$ over the frequencies $\w$ and $\w^{\prime}$. 
A long straightforward calculation then leads to the following expression for the velocity autocorrelation function
\begin{widetext}
\bea
\la \vv(t)\cdot \vv(0) \ra &=& \f{a^2 v_0^2}{4\pi} \int_{q_\mathrm{min}}^{q_\mathrm{max}} dq~q~\left[\f{\left[\left[(B+G)^2q^4(\W^2-D_r^2)+(\W^2+D_r^2)^2\right] \cos(\W~t) - 2\W D_r (B+G)^2 q^4 \sin(\W~t)\right]\text{e}^{-D_r t}}{\left[((B+G)^2q^4+\W^2-D_r^2)^2+4D_r^2\W^2 \right]} 
\right.\nonumber\\
&+&\left. \f{\left[\left[G^2q^4(\W^2-D_r^2)+(\W^2+D_r^2)^2\right] \cos(\W~t) - 2\W D_r G^2 q^4 \sin(\W~t)\right] \text{e}^{-D_r t}}{\left[(G^2q^4+\W^2-D_r^2)^2+4D_r^2\W^2 \right]} \right.\nonumber\\
&-&\left. \f{(B+G)q^2 D_r (D_r^2+\W^2-(B+G)^2q^4) \text{e}^{-(B+G)q^2 t}}{\left[(D_r^2+\W^2-(B+G)^2q^4)^2+4 (B+G)^2 q^4\W^2 \right]}-\f{G q^2 D_r (D_r^2+\W^2-G^2q^4) \text{e}^{-G q^2 t}}{\left[(D_r^2+\W^2-G^2 q^4)^2+4 G^2 q^4\W^2 \right]}\right].
\label{eq:velocity_autocorrelation_continuum}
\eea
\end{widetext}
We note that the integral in this expression must be computed numerically and that the integration limits $q_\mathrm{min}$ and $q_\mathrm{max}$ are determined by the largest and smallest scales in the system, respectively. Figure S3 in the Supplemental Material \cite{Supply2024} illustrates Eq.~(\ref{eq:velocity_autocorrelation_continuum}) in the limit $\Omega=0$ (panel a), the noiseless  limit $D_r=0$ (panel b), and for fixed $\Omega$ as a function of $D_r$ (panel c) and for a fixed $D_r$ as a function of $\Omega$. In all cases where $\Omega>D_r$, oscillations with frequency $\Omega$ appear.


\begin{figure}[!ht]
\includegraphics[width=8.5cm]{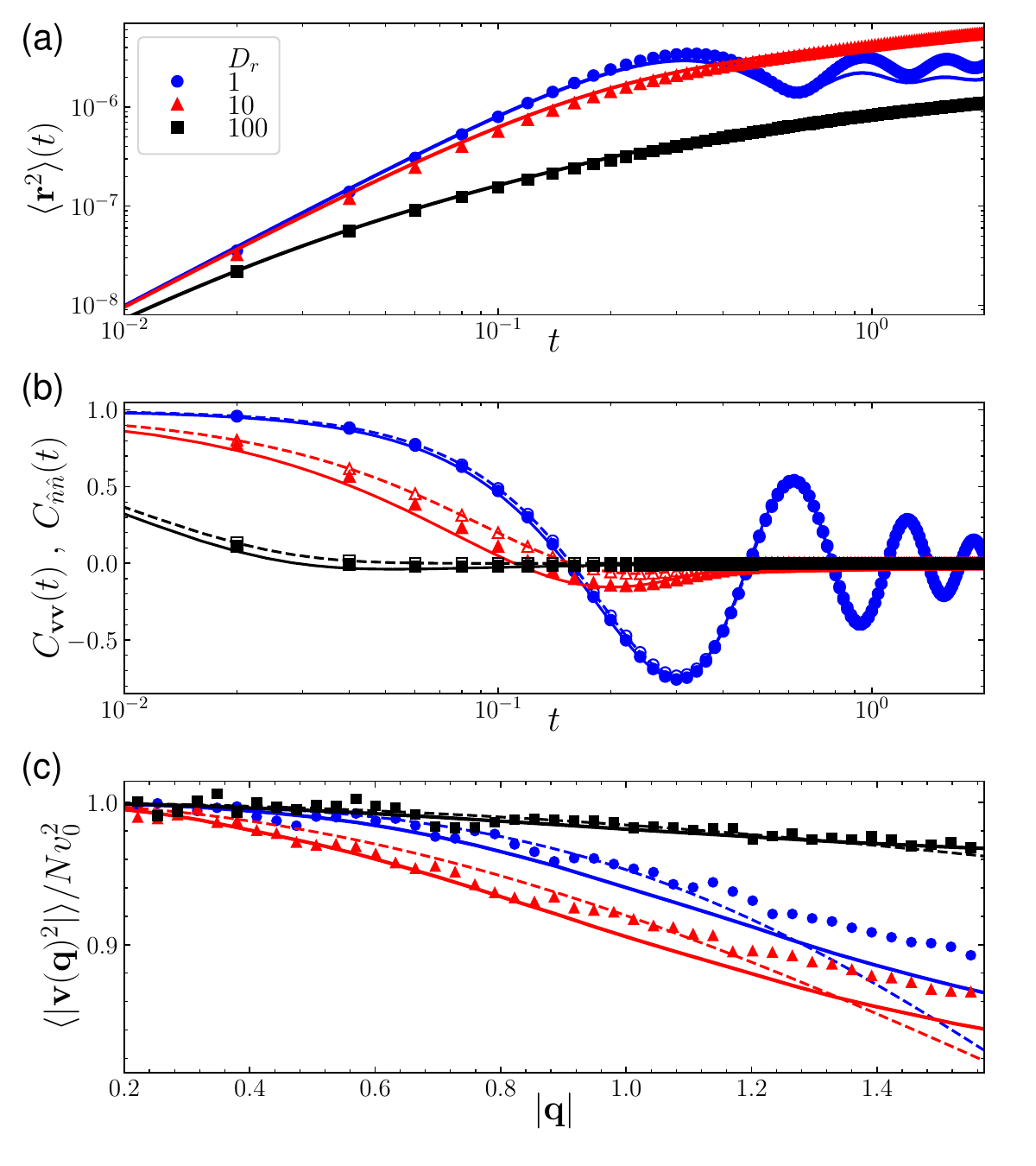}
\caption{Analysis of Non-monotonic Behavior at High Chirality and High Noise. (a) $\la\rv^2\ra(t)$ as function of time $t$.  The solid lines (Eq.~(\ref{eq:MSD})) results from single particle in a harmonic potential already explored in recent work~\cite{Debets2023}, detailed in Appendix-\ref{app:single_chiral_ht_MSD}. (b) $C_{\vv\vv}(t)=\la\vv(t)\cdot\vv(0)\ra/\la\vv(0)^2\ra$ (filled symbols and solid lines) and $C_{\nv\nv}(t)=\la\nv(t)\cdot\nv(0)\ra$ (open symbols and dashed lines) as a function of time $t$, solid lines results from the continuum elastic formulation Eq.~(\ref{eq:velocity_autocorrelation_continuum}) and dashed lines are the plot of Eq.~(\ref{eq:ncorr}). (c) $\la|\vv(\qv)^2|\ra/Nv_0^2$ as a function of $\qv$, solid lines obtain from normal mode formulation Eq.~(\ref{eq:spatial_vel_corr_mode}) and dashed lines are the results of continuum elastic formulation Eq.~(\ref{eq:spatial_vel_corr_continuum}). Symbols results from simulations in (a), (b), and (c) for $D_r=1$~(blue circles),~$10$~(red triangles),~$100$~(black squares) with $v_0=0.01$ and $\W=0.1$.}
\label{fig4}
\end{figure}

We finalize this section by exploring the scaling of the mean squared velocity in Eq.~(\ref{eq:spatial_vel_corr_continuum}) in two limiting cases: the achiral active limit for $\chi=1$ (setting $\W=0$) and the limit with no noise $\chi=0$ (setting $D_r=0$).
In the achiral active case, $\la|\vv(\qv)|^2\ra$ scales as $\sim(\xi_{\text{T}}q)^{-2}$ and in the no noise case, it scales as $\la|\vv(\qv)|^2\ra\sim(\xi_{\text{T}}q)^{-4}$.
In both cases, $\la|\vv(\qv)|^2\ra$ thus diverges for low $q$.

In these two limiting cases, we can also compute the closed-form analytic mean-squared velocity. 
In the achiral active limit~($\W=0$), we obtain the previously obtained result~\cite{Henkes2020}
\bea
\la |\vv|^2\ra  &=&  \frac{a^2 v_0^2}{8\pi} \left[ \frac{\log(1+\xi_{\text{L}}^2 q_{\mathrm{max}}^2)}{\xi_{\text{L}}^2} + \frac{\log(1+\xi_{\text{T}}^2 q_{\mathrm{max}}^2)}{\xi_{\text{T}}^2} \right].\nonumber
\label{eq:MSV_achiral}
\eea
This expression shows that the dominant scaling,
$\la |\vv|^2\ra\sim\xi_{\text{T}}^{-2}$ or $\la |\vv|^2\ra\sim\xi_{\text{L}}^{-2}$,
will follow the scaling Ansatz used for highly dense collective cellular motion in a monolayer~\cite{Garcia2015}. 
In the limit of no noise~($D_r=0$), we find
\bea
\la |\vv|^2\ra  &=& \frac{a^2 v_0^2}{8\pi} \left[ \frac{\tan^{-1}(\xi_{\text{L}}^2 q_{\mathrm{max}}^2)}{\xi_{\text{L}}^2}  + \frac{\tan^{-1}(\xi_{\text{T}}^2 q_{\mathrm{max}}^2)}{\xi_{\text{T}}^2} \right],\nonumber
\label{eq:MSV_CAPs}
\eea
which also shows the dominant scaling $\la |\vv|^2\ra\sim\xi_{\text{T}}^{-2}$ or $\la |\vv|^2\ra\sim\xi_{\text{L}}^{-2}$. 
Note that the mean-squared velocity will have small values in regimes of low chirality and low noise, which matches the regime where elastic energy is stored in the sheet, leading to the emergence of the mesoscopic correlated motion in the velocity fields.

\section{Comparison with simulations}
\label{sec_V}
To compare the analytical predictions developed in the previous section with simulations, we computed the spatial velocity correlations in Fourier space $\la|\vv(\qv)|^2\ra$ as well as the velocity and orientation autocorrelation functions ($C_{\vv\vv}(t)=\la \vv(t)\cdot \vv(0) \ra$ and $C_{\nv\nv}(t)=\la\nv(t)\cdot\nv(0)\ra$, respectively) for a broad range of values of $D_r$ and $\W$.
%
We focus on the two most salient regimes identified above, the emergence of correlated velocity fields for small $D_r$ and $\W$ values in Subsection~\ref{subsec:VA}, and on the extrema of the elastic and kinetic energy for high $D_r$ and $\W$ values in Subsection~\ref{subsec:VB}.

\subsection{Chiral Mesoscopic Range Order}
\label{subsec:VA}

Figure~\ref{fig3}(a) presents simulation snapshots of the velocity angles showing the emergence of a CMRO state displaying correlated velocity fields for small values of $D_r$ and $\W$, as shown in the analytically computed state space diagram in Fig.~\ref{fig1}(a).

Figures~\ref{fig3}(b) and \ref{fig3}(d) present the spatial velocity correlation in Fourier space, $\la|\vv(\qv)^2|\ra$ as a function of $|\qv|$, for a range of $D_r = 0.01, 0.1, 0.5, 1.0$ values with fixed $\W=0.1$, and for a range of $\W=0.01, 0.1, 0.5, 1.0$ values with fixed $D_r=0.01$.
We observe excellent agreement between the analytical normal mode formulation in Eq.~(\ref{eq:spatial_vel_corr_mode}) and the simulation results, represented respectively by solid lines and symbols, showing the emergence of correlated velocity fields for small $D_r$ and $\W$.
The continuum elastic formulation in Eq.~(\ref{eq:spatial_vel_corr_continuum}), displayed as dashed lines, shows good agreement with the simulations at low $q$, as expected.

Figures~\ref{fig3}(c) and \ref{fig3}(e), show the autocorrelation functions in time for the orientations and normalized velocities, labeled $C_{\nv\nv}(t)$ and $C_{\vv\vv}$, respectively. Here, the $C_{\nv\nv}(t)$ analytical curves in Eq.~(\ref{eq:ncorr}) are represented by dashed lines and their numerical values by open symbols, while the $C_{\vv\vv}$ analytical curves (\ref{eq:velocity_autocorrelation_continuum}) are displayed as solid lines and their numerical values as solid symbols. 
In order to best match the numerical integration in Eq.(\ref{eq:velocity_autocorrelation_continuum}) to our simulations, we chose the $q_\mathrm{min}$ and $q_\mathrm{max}$ that correspond to the smallest and largest simulated scales, as detailed in Appendix-\ref{app:continuum}.
The red curves and symbols in Fig.~\ref{fig3}(c) and Fig.~\ref{fig3}(e) show that both autocorrelation functions $C_{\nv\nv}(t)$ and $C_{\vv\vv}$ display oscillatory behavior for $D_r=0.01$ and $\W=0.1$.
The black curves and symbols show non-oscillatory behavior for higher nose $D_r=0.1$ in Fig.~\ref{fig3}(c) and for lower chirality $\W=0.01$ in Fig.~\ref{fig3}(e), displaying a faster decay in the $C_{\vv\vv}$ case.

\subsection{Elastic and Kinetic Energy Extrema}
\label{subsec:VB}
We now compare our analytical and numerical results on the elastic and kinetic energy extrema found in the high noise $D_r$, high chirality $\W$ regime.
First, we note that no such extrema is observed in active solids composed of achiral active particles, in which the mean stored elastic energy always decreases with noise $D_r$. This holds true for active solids composed of active particles with noisy chiral dynamics in the low chirality regime with $\W<\l_{max}$, as shown in Fig.~\ref{fig1}(a). 

In the high chirality regime $\W>\l_{max}$, however, there is a range of $D_r$ values for which the mean potential energy stored in all modes increases with $D_r$, as shown in Fig.~\ref{fig2}(e). This leads to a maximum in the mean stored elastic energy as a function of $D_r$(green dashed line in Fig.~\ref{fig1}(a)).
In the same regime, the mean-squared velocity obtained from the continuum elastic formulation displays non-monotonic behavior leading to a minimum in the kinetic energy (Fig.~\ref{fig2}(f) and magenta dashed line in Fig.~(\ref{fig1})(a)).

The non-monotonic features described above are also captured by the single particle case in an harmonic trap, corresponding to the curve with open black circles in Fig.~\ref{fig1}(a).
This can be seen in the Fig.~\ref{fig4}, which presents the mean square displacement $\la\rv^2\ra(t)$ in Fig.~\ref{fig4}(a),  $C_{\vv\vv}(t)$ in Fig.~\ref{fig4}(b), and $\la|\vv(\qv)|^2\ra$ in Fig.~\ref{fig4}(c), all in the high chirality regime ($\W=10$) and for noise strengths $D_r=1, 10, 100$.
We observe that, both, the long-time $\la\rv^2\ra$ values and the  $\la|\vv(\qv)|^2\ra$ curves exhibit non-monotonic behavior.
However, the single particle $C_{\vv\vv}$ result does not seem to capture this feature. 

These observations also allow us to return to the chiral glassy system of \cite{Debets2023}: 
The authors point out an emergent spatial correlation length in the limit of low $D_r$ and $\Omega$ while in the presence of oscillating velocity autocorrelations. We identify this with the CMRO state. Then other sets of states where the spatial but not the temporal correlations disappear as $\Omega$ increases correspond to the CD states, and finally the DD state when both disappear.

The same authors also identify a `hammering state' where particles oscillate in their cages, together with a maximum in the MSD and the diffusion constant. We can now identify this state and the peak in the diffusion constant with crossing the maximum of the energy spectrum along the $D_r$ direction in the CD state. Debets et al \cite{Debets2023} also present a version of the computation for a single particle in a harmonic trap.

\begin{figure}[!ht]
\includegraphics[width=8.5cm]{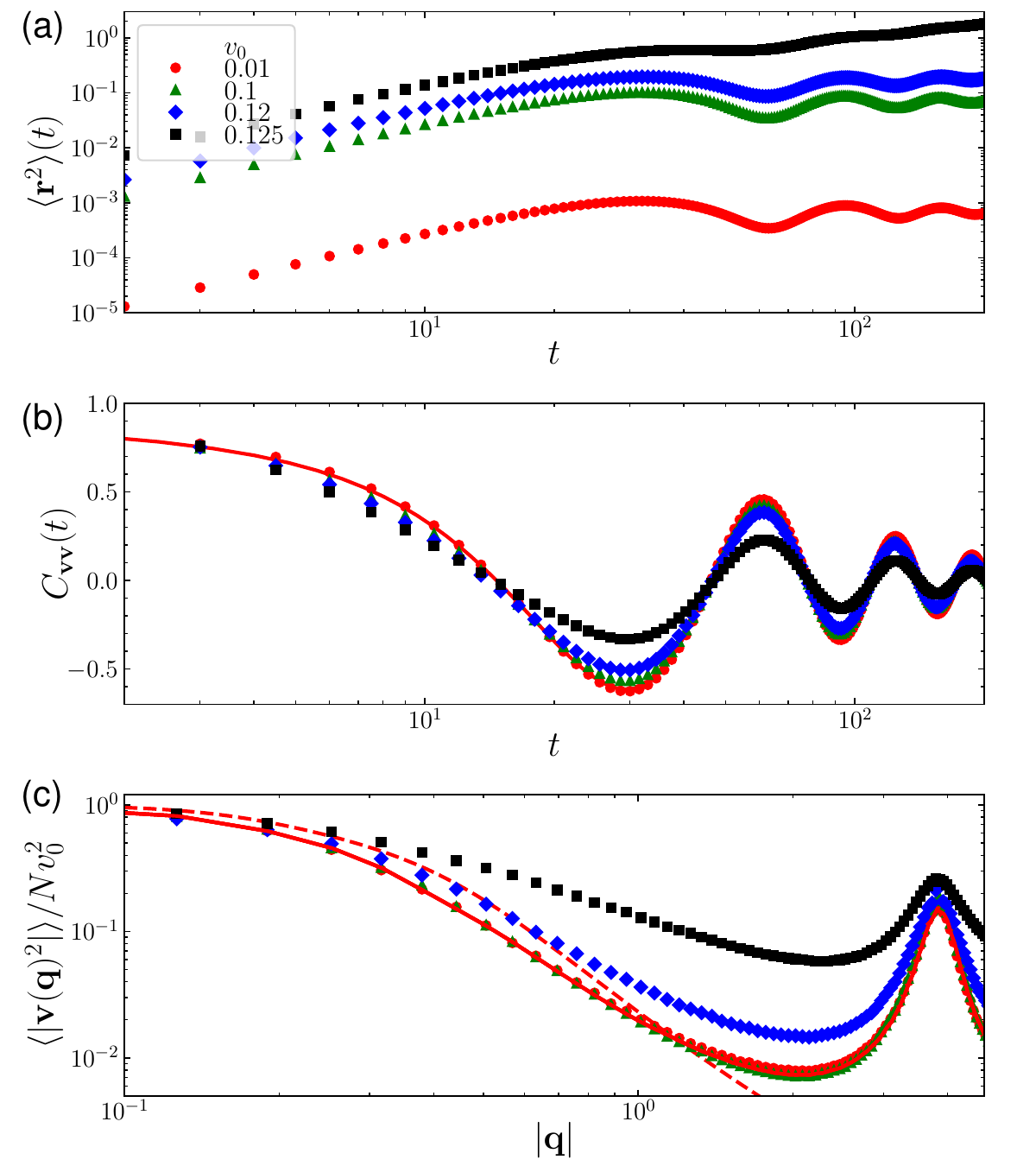}
\caption{Dynamics at Melting Regimes. Symbols results from simulation with varying active speed $v_0=0.01$~(red circles),~$0.1$~(green triangles),~$0.12$~(blue diamonds),~$0.125$~(black squares) with $D_r=0.01$ and $\W=0.1$. (a) $\la\rv^2\ra(t)$ as function of time $t$. (b) $C_{\vv\vv}(t)=\la\vv(t)\cdot\vv(0)\ra/\la\vv(0)^2\ra$ as a function of time $t$, solid line obtain from the continuum elastic formulation Eq.~(\ref{eq:velocity_autocorrelation_continuum}), compare with simulations (symbols). (c) $\la|\vv(\qv)^2|\ra/Nv_0^2$ as a function of $|\qv|$, solid line results of normal mode formulation Eq.~(\ref{eq:spatial_vel_corr_mode}) and dashed line results of continuum elastic formulation Eq.~(\ref{eq:spatial_vel_corr_continuum}), compare with simulations (symbols).
}
\label{fig5}
\end{figure}

\section{Analysis of Melting Behavior at High Activity}
\label{sec_VI}

The solid to liquid transition in systems of active particles has been extensively studied, starting with the observation that in the high density and low motility limit, active Brownian particles form crystals \cite{bialke2012crystallization} when monodisperse and glasses otherwise \cite{fily2014freezing}.

For two-dimensional equilibrium crystals 
this is a multifaceted process, distinguished by the progressive disintegration of both position and orientation coherence. 
In systems characterized by short-range interactions, melting manifests either as a first-order solid-liquid transition or via the sequential two-phase KTHNY mechanism involving solid-hexatic and hexatic-liquid transitions~\cite{Dash1999, Gasser2009, Li2020, Huang2020SoftMatter}.
Unlike passive systems, active crystals exhibit the capability to autonomously organize and transition into an active fluid state, facilitated by their intrinsic active speed and interaction forces. Though the melting transition in active particles to model the biological tissue occurs via a continuous solid-hexatic then followed by a continuous hexatic-liquid transition~\cite{digregorio2018full,Pasupalak2020}.

The glass transition in active Brownian particles has been the focus of an intensive research effort \cite{Berthier2014,Flenner2016,Berthier2017, Nandi2018,berthier2019glassy}, showing that the differences with the thermal glass transition are subtle, and that the transition is governed by an effective temperature $T_{\text{eff}}=v_0^2/2D_r$. This only changes in the limit $D_r\rightarrow0$ \cite{mandal2020extreme}, where the same mesoscopic length scale as discussed here becomes large \cite{Henkes2020, caprini2020hidden} and then starts to influence the transition properties \cite{berthier2019glassy,Flenner2020,keta24a}. Meanwhile, glasses of chiral active particles have to date only been thoroughly studied by Debets et al. \cite{Debets2023}; we discussed the applicability our results in the previous section.

%

In the high activity limit, our solid triangular monocrystal structure of chiral active particles with elastic interactions eventually melts. The details of this transition are beyond the scope of this investigation.
In Fig.~\ref{fig5}, we instead test the tolerance level of our analytic predictions in relation to increasing activity, specifically the active speed $v_0$. We consider both noise and chirality, which correspond to the chiral mesoscopic range order regimes, and incrementally increase $v_0$ to observe the melting behavior. The mean-squared displacement exhibits trapped oscillatory behavior until $v_0 = 0.12$, at which point melting begins, as shown in Fig.~\ref{fig5}(a). We also calculate the velocity autocorrelation functions (Fig.~\ref{fig5}(b)) and spatial velocity correlation functions in Fourier space (Fig.~\ref{fig5}(c)), and observe that the deviation starts at $v_0 > 0.1$.
%
The dynamics of velocity fields are shown in Supplemental Material Movie 6~\cite{Supply2024} for $v_0=0.1$ and in Supplemental Material Movie 7~\cite{Supply2024} for $v_0=0.12$ in CMRO regime with $D_r=0.01$ and $\W=0.1$. 
This clearly indicates that our analytic methods  
remain in excellent agreement with simulations until just below melting, but start deviating at high activity. Nonetheless, the temporal oscillations persist, and the mesoscale correlations persist, albeit with modified scaling. The latter is in agreement with what was observed when fitting the non-chiral models to cell sheet data \cite{Henkes2020} and using simulations of active Brownian particles (ABPs) at higher activity \cite{Henkes2020,keta24a}.
This also explains why the results here are predictive for the active glassy dynamics investigated in Debets et al. \cite{Debets2023}.

\begin{figure*}[!ht]
\includegraphics[width=8.0cm]{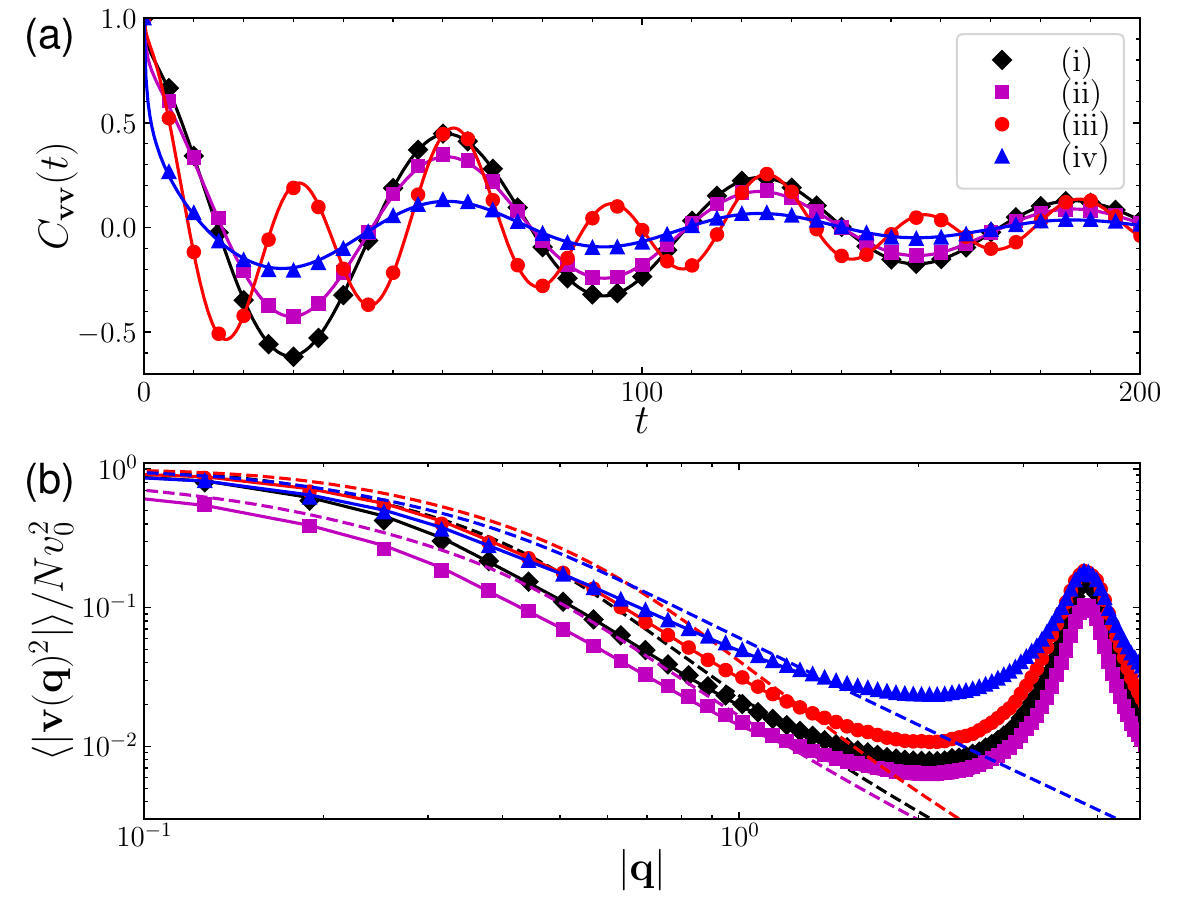}
\includegraphics[width=8.0cm]{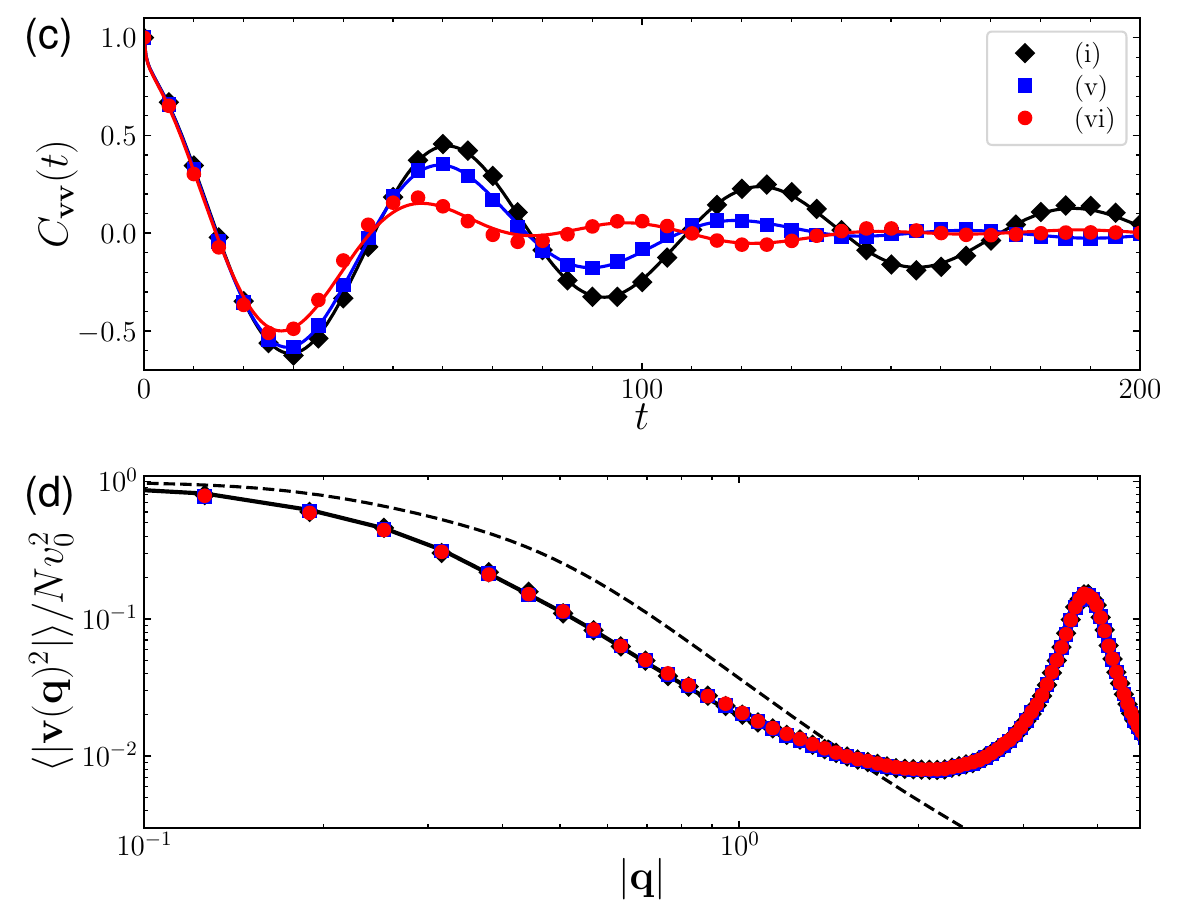}
\caption{Dynamics of Heterogeneous Active Solids. (a),(b) Active solids composed of binary mixtures with packing fractions $\phi_A=\phi_B=0.5$. (i) $D_{r}^{A}=D_{r}^{B}=0.01$ and $\W_A=\W_B=0.1$. (ii) $D_{r}^{A}=D_{r}^{B}=0.01$, $\W_A=0.0$, and $\W_B=0.1$. (iii) $D_{r}^{A}=D_{r}^{B}=0.01$, $\W_A=0.1$, and $\W_B=0.2$. (iv) $D_{r}^{A}=0.01$, $D_{r}^{B}=0.1$, and $\W_A=\W_B=0.1$. (c),(d) Active solids composed of heterogeneous particles with uniform distributions of ($D_r,\W$) in (v) $\pm 20\%$ and (vi) $\pm 40\%$ from uniform mixtures in (i). (a),(c) $C_{\vv\vv}(t)=\la\vv(t)\cdot\vv(0)\ra/\la\vv(0)^2\ra$ as a function of time $t$, obtain from continuum elastic formulation (solid lines), compare with simulations (symbols). (b),(d) $\la|\vv(\qv)|^2\ra/Nv_0^2$ as a function of $|\qv|$, result from the continuum elastic formulation (dashed lines) and normal mode formulation (solid lines), compare with simulations (symbols).
}
\label{fig6}
\end{figure*}    
\section{Extension to Heterogeneous Systems}
\label{sec_VII}
We now extend our results to heterogeneous systems of active particles, where they do not all have the same dynamical properties.
We will first study binary mixtures of particles with different chiralities and rotational diffusion constants in Subsection~\ref{subsec_binary_mixtures}, and then analyze cases where the particles have a distribution of chirality and diffusion values in Subsection~\ref{subsec_heterogeneous_particles}.

\subsection{Binary Mixtures}
\label{subsec_binary_mixtures}
We consider active solids composed of two particle species, $A$ and $B$, differentiated by their $D_r$ and $\W$ values. Each represents a fraction of the total, corresponding to their respective packing fractions, $\phi_A$ and $\phi_B$. Since the active driving acts as a time-correlated but single-particle noise (see Eq.~(\ref{eq:noise_mode}) and Eq.~(\ref{eq:force_corr_continuum})), the total driving noise from two species can be computed by simple superposition, with no particle-particle cross-correlations.
Then quantities such as $\la|\vv(\qv)|^2\ra$ and $\la \vv(t)\cdot \vv(0) \ra$, that we analytically derived in Sec.~\ref{sec_IV} can be expressed as the average of the $A$ and $B$ contributions, weighted by their respective fractions of the total.
In the linear response regime, the general expression for 
$f = |\vv(\qv)|^2$ or $f = \vv(t)\cdot \vv(0)$ 
is thus given by the following superposition formula
\bea
\label{eq:BinarySuper}
\la f\ra &=& \f{\phi_{A}}{\phi} \la f\ra_{A} + \f{(\phi-\phi_{A})}{\phi} \la f\ra_{B}  ~,
\eea
where $\la \cdot \ra_{A}$ and $\la \cdot \ra_{B}$ represent the mean over populations $A$ or $B$, respectively, and $\phi$ is the total packing fraction.
We note that a mixture of positive and negative chiral active particles, with the same absolute chirality but different sign, behaves just as a system with uniform chirality due to the symmetry $\la f\ra_{A}(\W)=\la f\ra_{B}(-\W)$.
Despite this, the CMRO state shows the emergence of patches of uniform positive and negative chirality, in a proportion controlled by the ratio of $\phi_{A}/\phi_{B}$ (see Supplemental Material Movie 8~\cite{Supply2024}).
We also note that in the case of a random mixture of chiral active and passive particles, where $v_0^{\mathrm{A}} > 0$ and $v_0^{\mathrm{B}} = 0$, the dynamics of the CMRO state is controlled by $\phi_A/\phi$.

To illustrate the effects of having binary mixtures of particles, we present simulations and analytical results for three different species combinations, all displaying chiral mesoscopic range order.  
Figures~\ref{fig6}(a),(b) show the 
$C_{\vv\vv}(t)$
and
$\la|\vv(\qv)|^2\ra/Nv_0^2$
obtained for three different binary mixtures with equal fractions $\phi_A/\phi_B=1$ and total packing fraction $\phi=1$, in addition to the single species, uniform case displayed as the curves labeled (i). 
First, we consider mixtures of achiral ($\W_A=0$) and chiral  ($\W_B=0.1$) active particles, with equal rotational diffusion coefficients $D_r^A=D_r^B=0.01$, displayed as the (ii) curves.
Second, we consider mixtures of particles with two different chirality values ($\W_A=0.1$ and $\W_B=0.2$) and the same $D_r^A=D_r^B=0.01$, displayed as the (iii) curves (see Supplemental Material Movie 9~\cite{Supply2024}).
Moreover, we consider mixtures of particles with equal chirality values $\W_A=\W_B=0.1$ and two different rotational diffusion coefficients, $D_r^A=0.01$ and $D_r^B=0.1$, displayed as the (iv) curves.

Figure \ref{fig6}(a) shows that the clear oscillations in the velocity autocorrelation function are suppressed when considering achiral-chiral mixtures (ii), compared to the single species case (i). 
It also shows that the velocity autocorrelation displays oscillations with both periods, $2\pi/\W_A$ and $2\pi/\W_B$, in systems with two different chirality values (iii).
Finally, it shows that mixtures of chiral active particles with two different rotational diffusion coefficients (iv) suppress this oscillatory behavior following the lower persistence length of the population with the highest $D_r$.
Figure \ref{fig6}(b) shows that binary mixtures of achiral and chiral active particles (ii) enhance the ordering in the CMRO state when compared to uniform mixtures (i), decaying at lower $|\qv|$ values.
On the other hand, chiral active binary mixtures with, both, different chirality values (iii) and different rotational diffusion coefficients (iv) suppress the ordering in the CMRO state when compared to uniform mixtures (i), decaying at larger $|\qv|$ values.
Finally, we note that the analytic superposition results (solid and dashed lines) show excellent agreement with the simulation results (symbols) for all curves in Fig.~\ref{fig6}(a,b), thus validating Eq.~(\ref{eq:BinarySuper}).

Figure S4 in the Supplementary Information \cite{Supply2024} further illustrates how oscillations appear, disappear and interfere using plots of the velocity time correlations computed using Eq.~(\ref{eq:BinarySuper}) with sliding scale binary mixtures.

\subsection{Heterogeneous Particles}
\label{subsec_heterogeneous_particles}
We now further extend our analyses to investigate heterogeneous active solids composed of particles that follow a distribution of dynamical properties.
To do this, we consider a complex mixture of $n$ different types of active particles, each with corresponding packing fraction $\phi_1,\phi_2,...,\phi_n$, adding up to a total packing fraction $\phi=\sum_{i=1}^{n}\phi_i$. 
We can then obtain a general expression for the combined values 
$f = |\vv(\qv)|^2$ and $f = \vv(t)\cdot \vv(0)$, in terms of the individual components as
\bea
\label{eq:MultiSuper}
\la f\ra =\sum_{i=1}^{n} \phi_i \la f\ra_i /\phi .
\eea
where $\la \cdot \ra_{i}$ represents the mean over population type $i$.
With this general superposition expression, we can analytically predict the dynamics for active solids composed of diverse combinations of active particles, in the linear response regime.

Figures \ref{fig6}(c),(d), for example, presents the analytical and simulated spatiotemporal correlations for the active solid dynamics of systems of particles with a uniformly distributed range of $D_r$ and $\W$, spanning (i) $\pm 0\%$, (v) $\pm 20\%$, and (vi) $\pm 40\%$ of their mean value $(D_r,\W)=(0.01,0.1)$.
Figure \ref{fig6}(c) shows a clear deviation of the velocity-velocity correlations $C_{\vv\vv}$ in the heterogeneous cases (v) and (vi) with respect to the homogeneous case (i). 
We note that the analytic superposition of Eq.~(\ref{eq:velocity_autocorrelation_continuum}) using Eq.~(\ref{eq:MultiSuper}), displayed as solid lines, perfectly captures the simulation results, represented by symbols.
Figure \ref{fig6}(d) shows that $\la|\vv(\qv)|^2\ra/Nv_0^2$ is practically invariant under variations of the parameters $D_r$ and $\W$.
We also note that the analytic superposition of Eq.~(\ref{eq:spatial_vel_corr_mode}) and Eq.~(\ref{eq:spatial_vel_corr_continuum}), respectively represented by a solid and a dashed line, can properly capture the simulation results, displayed as symbols.
The figure shows that the spatial velocity correlations are insensitive to constructing heterogeneous systems but that the temporal velocity autocorrelation is affected by heterogeneity. 
Indeed, increased heterogeneity in particle chirality leads to desynchronization of the oscillations, while heterogeneity in $D_r$ does not (shown in Fig. S5a and S5c in the Supplemental Material~\cite{Supply2024}). In neither case are the spatial correlations affected (Fig. S5b and S5d).
This predicts that oscillatory dynamics will only be apparent in systems with a relatively homogeneous level of chirality of the active particles.

\section{Conclusions}
\label{sec_VIII}

In this work, we formulated the analytic linear response theory for an active solid composed of self-propelled particles with noisy chiral dynamics. We considered a minimal model with the potential for describing a broad range of systems, ranging from artificial active solids made of chiral self-propelled robots to biological tissues with emergent macroscopic chiral order. We developed an analytic formulation that allowed us to fully describe all the observed dynamics in the linear response regime, perfectly matching our numerical results.

We described the emergence of four different regimes in the phase space formed by the chirality $\W$ and the rotational diffusion coefficient $D_r$. For small enough $D_r$ and $\W$, we observe chiral (CMRO) and achiral (MRO) self-organized states displaying mesoscopic range order. 
For larger $D_r$ and $\W$, we only find chiral (CD) and achiral (DD) disordered states.
In all cases, the chiral sates appear for $D_r<\W$ and the achiral states, for $D_r>\W$.
We then explored the dynamics of the melting regime by increasing the activity $v_0$, showing that our analytic results for the spatial velocity correlations and the temporal velocity autocorrelations agree with simulations up to an active speed $v_0 \approx 0.1$, just below $v_0=0.12$, the melting point predicted by our active solid theory.
Finally, we showed that our analytic approaches can be extended to consider particles with heterogeneous dynamical features, including different chirality and rotational diffusion levels. We derived analytic superposition expressions for binary and more complex mixtures of heterogeneous active particles, demonstrating their excellent agreement with simulations. 

Our work is consistent with the (sparse) literature on chiral ABP solids. Notably we recover the oscillatory correlations and `hammering' resonance observed by \cite{Debets2023} in the glassy state.
Recently, Caprini et al.~\cite{caprini2024self} explored the emergence of self-reverting vortices in the absence of alignment interactions as a result of the interplay between attractive interactions and chirality. They observed two kinds of vortices with persistent and oscillatory behavior, similar to our observation that the correlated velocity fields in the mesoscopic range can either be persistent (MRO) or oscillatory (CMRO). 
However, due to subtly different equations that include inertia and spatial noise, we cannot currently compare results quantitatively.

We hope that our analytical descriptions of dense, solid chiral active systems can help establish a foundation for the systematic understanding of the emergent dynamics in this type of systems. Future research could explore the scalability of our theory, its practical applications in synthetic biology, and its potential impact on the design and control of rotating correlated velocity fields in active solids.

\section*{Acknowledgements}
This project was made possible through the support of Grant 62213 from the John Templeton Foundation. SH would like to acknowledge support of the NWO through her Leiden University startup package within the sector plan.
\vspace{1cm}

\appendix

\section{Orientation Autocorrelation}
\label{app:orientation_autocorr}
We adopt the method of exact moments calculation for stiff chains, as described in Hermans et al.~\cite{Hermans1952}, to determine the exact moments for active dynamics. This approach, previously studied in \cite{Shee2020ActiveMoments, Chaudhuri2021, Shee2022}, is utilized to obtain the exact dynamics for a chiral active Brownian particle in a harmonic trap, as also detailed in Debets et al.~\cite{Debets2023}. Utilizing the Laplace transform $\tilde P(\nv, s) = \int_0^\infty dt\, \text{e}^{-s t}\, P(\nv, t) $ in Eq.~(\ref{F-P-orientation}) and defining the mean of an observable $\la \psi \ra_s = \int d\nv\, \psi(\nv ) \tilde P(\nv, s)$, multiplying by $\psi(\nv)$ and integrating over all possible $\nv$,  we find
\bea
-\la \psi \ra_0 + s \la \psi \ra_s &=& D_r \la \nabla_\nv^2 \psi \ra_s + \W~ \la \nv^{\perp}\cdot \nabla_{\nv} \psi\ra_s~,
\label{moment}
\eea
where the initial condition sets $\la \psi \ra_0 = \int d\nv\,  \psi(\nv) P(\nv, 0)$. Without any loss of generality, we consider the initial condition to follow $P(\nv, 0) = \d(\nv - \nv_0)$, where $\nv_0$ is the initial orientation of the particle. 

We utilize Eq.~(\ref{moment}) to compute exact orientation correlation as a function of time.
We consider the initial orientation of the particles along $\la\nv\ra=\nv_0$. To calculate $\la\nv\ra$, we use $\psi =\nv$ in the Eq.~(\ref{moment}), leads to
\bea
\la\nv\ra_s &=& \frac{\nv_0 + \W~ \la\nv^{\perp}\ra_s}{s+D_{r}}~,
\label{eq:nav_Laplace}
\eea
where we can calculate $\la\nv^{\perp}\ra_s$ with initial perpendicular orientation $\la\nv^{\perp}\ra=\nv^{\perp}_0$, using $\psi=\nv^{\perp}$ in the Eq.~(\ref{moment}) gives $\la\nv^{\perp}\ra_s = (\nv^{\perp}_0 - \W \la\nv\ra_s)/(s+D_r)$, substituting back into the Eq.~(\ref{eq:nav_Laplace})
\bea
\la\nv\ra_s &=& \frac{\nv_0(s+D_r)+ \W~ \nv^{\perp}_0}{(s+D_r)^2+\W^2}~.
\eea
Inverse Laplace transform leads to $\la\nv(t)\ra = \text{e}^{-D_r t} [\nv_0 \cos(\W t)+ \nv^{\perp}_0 \sin(\W t)]$. Taking the dot product with the initial orientation, we get the orientation autocorrelation in Eq.~(\ref{eq:ncorr}).

\section{Single Chiral Active Brownian Particle in a Harmonic Trap}
\label{app:single_chiral_ht_MSD}

Following  a similar procedure of orientation autocorrelation calculation, we can also calculate MSD $\la\rv^2\ra$ of single chiral active Brownian particle (CABP) in a harmonic trap, as already explored in~\cite{Debets2023}
\bea
&&\la\rv^2\ra(t) = \f{v_0^2(D_r+\mu k)}{\mu k [(D_r+\mu k)^2+\W^2]} - \f{v_0^2(D_r-\mu k)\text{e}^{-2\mu k t}}{\mu k [(D_r-\mu k)^2+\W^2]}\nonumber\\
&&+\f{2 v_0^2\text{e}^{-(D_r+\mu k) t}[( D_r^2-\mu^2k^2-\W^2)\cos(\W t)-2D_r\W \sin(\W t)]}{ [(D_r-\mu k)^2+\W^2][(D_r+\mu k)^2+\W^2]}\nonumber\\
\label{eq:MSD}
\eea
We plot the above equation in Fig.~\ref{fig4}(a) to compare the MSD of dense chiral active systems in high $D_r$ and high $\W$ regimes, demonstrating that their behavior is similar to that of a single CABP in a harmonic trap.
Simplifying the above Eq.~(\ref{eq:MSD}) in the limit of $D_r t << 1$ and $\W t<<1$ yields
\bea
\la\rv^2\ra(t) &=& v_0^2 t^2 - \f{v_0^2}{3}(D_r +3\mu k)  t^3 + \f{v_0^2}{12}(D_r^2 + 4D_r \mu k \nonumber\\
&&+ 7 \mu^2 k^2-\W^2)t^4+ \mathcal{O}(t^5)
\eea
In the long time limit, we get the steady-state MSD, $\la\rv^2\ra_{st} = \la\rv^2\ra(t)|_{t\to\infty}$,
\bea
\la\rv^2\ra_{st} &=& \f{v_0^2(D_r+\mu k)}{\mu k[(D_r+\mu k)^2+\W^2]}
\label{eq:disp_fluct_long_time}
\eea
The value of the steady-state MSD, $\la\rv^2\ra_{st}$ reaches its maximum for a constant $\W$ along the line of $D_r$ when $D_r^{*}=\W-\mu k$(see the black dotted line on the $D_r-\W$ plane in Fig.~\ref{fig1}(a) with $\mu=1$ and $k=1$). It suggests that the re-entrant transition from low $\la\rv^2\ra_{st}$ to high $\la\rv^2\ra_{st}$ (which reaches maximum at $D_r^{*}$) and then back to low $\la\rv^2\ra_{st}$ with an increase in $D_r$ occurs only when chirality is high i.e., $\W>\mu k$.

\section{Determining the Elastic Moduli and Continuum Theory Estimation to Compare with Simulations}
\label{app:continuum}
We first equilibrate the non-equilibrium steady state configurations by letting $v_0=0$ in Eq.~(\ref{eom1}). The positions reaches to its equilibrium positions at long time. Utilizing the equilibrated positions, we construct dynamical Hessian matrix. We transform this dynamical matrix to the Fourier space with appropriate grid space of $\qv$. The longitudinal and transverse eigenvalues are $(B+G) q^2$ and $Gq^2$, respectively.

We determine the bulk modulus $B$ and shear modulus $G$ by performing a linear fit of the radially averaged longitudinal and transverse eigenvalues against $q^2$, focusing on the limit where $q^2\leq 1$. Our calculations for soft disks yield the moduli values:~:~$G=0.61$, $B=2.03$, and $(B+G)/G=4.33$ with relative error $<1\%$ (averaging over 10 independent estimations).


When we compare numerical results with the continuum theory, the simulations are done for relatively large but finite systems of size $L$, with minimum length scale given by the particle size $a=2r_0$, where $r_0$ is the radius of the particle. 
This guide us to perform discrete space Fourier transformation in numerical analysis. On the other hand, by setting $L\to \infty$ and $a\to 0$ in the analytic calculation give the results in hydrodynamic limit. 

For consistency between two approaches, we use the following space continuous Fourier transform
\bea
\uv (\rv,t) =  \frac{1}{(2\pi)^2} \int d^2\qv~\tilde{\uv}(\qv,t) \text{e}^{-i\qv\cdot\rv}.
\eea
By considering the finite system and particle sizes, we discretize the integral into
$$\frac{1}{(2\pi)^2} \int d^2\qv \to \frac{1}{Na^2}  \sum_{\qv} ~~\text{and}~~ \int d^2\rv \to a^2 \sum_{r},$$
where $N=4\phi L^2/\pi a^2$, $\phi$ is the packing fraction of the system close to $1$ for dense systems. In the sum $\qv$ takes discrete values defined by the geometry of the problem. For instance a square lattice of linear size $L$, $\qv\equiv(q_x,q_y)=2\pi/L(m,n)$ where integers $m,n$ satisfying $0\leq m,n\leq L/a-1$. Thus, the discrete space Fourier transform $\uv(\qv,t)$ is related to the continuous Fourier transform $\tilde{\uv}(\qv,t)$ through $\tilde{\uv} (\qv,t) = a^2 \uv(\qv,t)$.

\bibliographystyle{prsty}
\bibliography{reference}

\begin{thebibliography}{100}

\bibitem{Kummel2013}
F. K{\"{u}}mmel, B. ten Hagen, R. Wittkowski, I. Buttinoni, R. Eichhorn, G. Volpe, H. L{\"{o}}wen, and C. Bechinger, Physical Review Letters {\bf 110},  198302  (2013).

\bibitem{Bechinger2016}
C. Bechinger, R. Di~Leonardo, H. L{\"{o}}wen, C. Reichhardt, G. Volpe, and G. Volpe, Reviews of Modern Physics {\bf 88},  45006  (2016).

\bibitem{Mano2017}
T. Mano, J.~B. Delfau, J. Iwasawa, and M. Sano, Proceedings of the National Academy of Sciences of the United States of America {\bf 114},  E2580  (2017).

\bibitem{Zhang2022}
B. Zhang and A. Snezhko, Physical Review Letters {\bf 128},  218002  (2022).

\bibitem{Grzybowski2000}
B.~A. Grzybowski, H.~A. Stone, and G.~M. Whitesides, Nature 2000 405:6790 {\bf 405},  1033  (2000).

\bibitem{Cruz2024}
J.-M. Cruz, O. Díaz-Hernández, A. Castañeda-Jonapá, G. Morales-Padrón, A. Estudillo, and R. Salgado-García, Soft Matter {\bf 20},  1199  (2024).

\bibitem{Jennings1901}
H.~S. Jennings, The American Naturalist {\bf 35},  369  (1901).

\bibitem{Loose2014}
M. Loose and T.~J. Mitchison, Nature Cell Biology {\bf 16},  38  (2014).

\bibitem{Sumino2012}
Y. Sumino, K.~H. Nagai, Y. Shitaka, D. Tanaka, K. Yoshikawa, H. Chat{\'{e}}, and K. Oiwa, Nature {\bf 483},  448  (2012).

\bibitem{Brokaw1982}
C.~J. Brokaw, D.~J. Luck, and B. Huang, Journal of Cell Biology {\bf 92},  722  (1982).

\bibitem{DiLuzio2005}
W.~R. DiLuzio, L. Turner, M. Mayer, P. Garstecki, D.~B. Weibel, H.~C. Berg, and G.~M. Whitesides, Nature {\bf 435},  1271  (2005).

\bibitem{Lauga2006}
E. Lauga, W.~R. DiLuzio, G.~M. Whitesides, and H.~A. Stone, Biophysical Journal {\bf 90},  400  (2006).

\bibitem{Riedel2005}
I.~H. Riedel, K. Kruse, and J. Howard, Science {\bf 309},  300  (2005).

\bibitem{Nosrati2015}
R. Nosrati, A. Driouchi, C.~M. Yip, and D. Sinton, Nature Communications {\bf 6},  8703  (2015).

\bibitem{Van2008}
S. van Teeffelen and H. L\"owen, Phys. Rev. E {\bf 78},  020101  (2008).

\bibitem{Van-Teeffelen2009}
S. Van~Teeffelen, U. Zimmermann, and H. L{\"{o}}wen, Soft Matter {\bf 5},  4510  (2009).

\bibitem{Mijalkov2013}
M. Mijalkov and G. Volpe, Soft Matter {\bf 9},  6376  (2013).

\bibitem{Volpe2014}
G. Volpe, S. Gigan, and G. Volpe, American Journal of Physics {\bf 82},  659  (2014).

\bibitem{Lowen2016}
H. L{\"{o}}wen, The European Physical Journal Special Topics {\bf 225},  2319  (2016).

\bibitem{Sevilla2016}
F.~J. Sevilla, Phys. Rev. E {\bf 94},  062120  (2016).

\bibitem{Morelly2019}
S. Morelly, M. Han-Mei~Tang, N.~J. Alvarez, S. Muhuri, M. Rao, S. Ramaswamy, T. Markovich, E. Tjhung, and M.~E. Cates, New Journal of Physics {\bf 21},  112001  (2019).

\bibitem{Chepizhko2020}
O. Chepizhko and T. Franosch, New Journal of Physics {\bf 22},  073022  (2020).

\bibitem{Caprini2023}
L. Caprini, H. L{\"{o}}wen, and U. Marini Bettolo~Marconi, Soft Matter {\bf 19},  6234  (2023).

\bibitem{Leonardo2011}
R. Di~Leonardo, D. Dell’Arciprete, L. Angelani, and V. Iebba, Physical Review Letters {\bf 106},  038101  (2011).

\bibitem{Araujo2019}
G. Araujo, W. Chen, S. Mani, and J.~X. Tang, Biophysical Journal {\bf 117},  346  (2019).

\bibitem{Bohmer2005}
M. B{\"{o}}hmer, Q. Van, I. Weyand, V. Hagen, M. Beyermann, M. Matsumoto, M. Hoshi, E. Hildebrand, and U.~B. Kaupp, The EMBO Journal {\bf 24},  2741  (2005).

\bibitem{Taktikos2011}
J. Taktikos, V. Zaburdaev, and H. Stark, Physical Review E {\bf 84},  41924  (2011).

\bibitem{Gibbs2009}
J.~G. Gibbs and Y. Zhao, Small {\bf 5},  2304  (2009).

\bibitem{Gibbs2011}
J.~G. Gibbs, S. Kothari, D. Saintillan, and Y.-P. Zhao, Nano letters {\bf 11},  2543  (2011).

\bibitem{Denk2016}
J. Denk, L. Huber, E. Reithmann, and E. Frey, Physical Review Letters {\bf 116},  178301  (2016).

\bibitem{Bar2020}
M. B{\"{a}}r, R. Gro{\ss}mann, S. Heidenreich, and F. Peruani, Annual Review of Condensed Matter Physics {\bf 11},  441  (2020).

\bibitem{Zhang2020}
B. Zhang, A. Sokolov, and A. Snezhko, Nature Communications {\bf 11},  4401  (2020).

\bibitem{Campbell2017}
A.~I. Campbell, R. Wittkowski, B. Ten~Hagen, H. L{\"{o}}wen, and S.~J. Ebbens, Journal of Chemical Physics {\bf 147},  84905  (2017).

\bibitem{Archer2015}
R.~J. Archer, A.~I. Campbell, and S.~J. Ebbens, Soft Matter {\bf 11},  6872  (2015).

\bibitem{Lei2023}
T. Lei, C. Zhao, R. Yan, and N. Zhao, Soft Matter {\bf 19},  1312  (2023).

\bibitem{Liebchen2017}
B. Liebchen and D. Levis, Physical Review Letters {\bf 119},  058002  (2017).

\bibitem{Levis2018}
D. Levis and B. Liebchen, Journal of Physics: Condensed Matter {\bf 30},  084001  (2018).

\bibitem{Liao2018}
G.~J. Liao and S.~H. Klapp, Soft Matter {\bf 14},  7873  (2018).

\bibitem{Kaiser2013}
A. Kaiser and H. L{\"{o}}wen, Physical Review E {\bf 87},  032712  (2013).

\bibitem{Liu2019}
Y. Liu, Y. Yang, B. Li, and X.-Q. Feng, Soft Matter {\bf 15},  2999  (2019).

\bibitem{Ma2022}
Z. Ma and R. Ni, The Journal of Chemical Physics {\bf 156},  21102  (2022).

\bibitem{Semwal2024}
V. Semwal, J. Joshi, and S. Mishra, Physica A: Statistical Mechanics and its Applications {\bf 634},  129435  (2024).

\bibitem{Bickmann2022}
J. Bickmann, S. Br{\"{o}}ker, J. Jeggle, and R. Wittkowski, The Journal of Chemical Physics {\bf 156},  194904  (2022).

\bibitem{Lei2019}
Q.-L. Lei and R. Ni, Proceedings of the National Academy of Sciences {\bf 116},  22983  (2019).

\bibitem{Lei2019Science}
Q.-L. Lei, M.~P. Ciamarra, and R. Ni, Science Advances {\bf 5},  eaau7423  (2019).

\bibitem{Hrishikesh2023}
B. Hrishikesh and E. Mani, Soft Matter {\bf 19},  225  (2023).

\bibitem{Levis2019PRE}
D. Levis and B. Liebchen, Physical Review E {\bf 100},  012406  (2019).

\bibitem{Ceron2023}
S. Ceron, K. O’Keeffe, and K. Petersen, Nature Communications {\bf 14},  940  (2023).

\bibitem{bililign2022motile}
E.~S. Bililign, F. Balboa~Usabiaga, Y.~A. Ganan, A. Poncet, V. Soni, S. Magkiriadou, M.~J. Shelley, D. Bartolo, and W.~T. Irvine, Nature Physics {\bf 18},  212  (2022).

\bibitem{Tan2022}
T.~H. Tan, A. Mietke, J. Li, Y. Chen, H. Higinbotham, P.~J. Foster, S. Gokhale, J. Dunkel, and N. Fakhri, Nature {\bf 607},  287  (2022).

\bibitem{Drescher2009}
K. Drescher, K.~C. Leptos, I. Tuval, T. Ishikawa, T.~J. Pedley, and R.~E. Goldstein, Physical Review Letters {\bf 102},  168101  (2009).

\bibitem{Petroff2015}
A.~P. Petroff, X.-L. Wu, and A. Libchaber, Physical Review Letters {\bf 114},  158102  (2015).

\bibitem{Yan2015}
J. Yan, S.~C. Bae, and S. Granick, Soft Matter {\bf 11},  147  (2015).

\bibitem{Huang2020SoftMatter}
Z.-F. Huang, A.~M. Menzel, and H. L\"owen, Phys. Rev. Lett. {\bf 125},  218002  (2020).

\bibitem{Ishikawa2020}
T. Ishikawa, T.~J. Pedley, K. Drescher, and R.~E. Goldstein, Journal of Fluid Mechanics {\bf 903},  A11  (2020).

\bibitem{Petroff2023}
A.~P. Petroff, C. Whittington, and A. Kudrolli, Physical Review E {\bf 108},  014609  (2023).

\bibitem{fruchart2023odd}
M. Fruchart, C. Scheibner, and V. Vitelli, Annual Review of Condensed Matter Physics {\bf 14},  471  (2023).

\bibitem{Henkes2011}
S. Henkes, Y. Fily, and M.~C. Marchetti, Physical Review E {\bf 84},  040301(R)  (2011).

\bibitem{Lin2021}
G. Lin, Z. Han, and C. Huepe, New Journal of Physics {\bf 23},  023019  (2021).

\bibitem{Baconnier2022}
P. Baconnier, D. Shohat, C.~H. L{\'{o}}pez, C. Coulais, V. D{\'{e}}mery, G. D{\"{u}}ring, and O. Dauchot, Nature Physics {\bf 18},  1234  (2022).

\bibitem{Lin2023}
G. Lin, Z. Han, A. Shee, and C. Huepe, Physical Review Letters {\bf 131},  168301  (2023).

\bibitem{Baconnier2023}
P. Baconnier, D. Shohat, and O. Dauchot, Physical Review Letters {\bf 130},  028201  (2023).

\bibitem{Xu2023}
H. Xu, Y. Huang, R. Zhang, and Y. Wu, Nature Physics {\bf 19},  46  (2023).

\bibitem{Berthier2013}
L. Berthier and J. Kurchan, Nature Physics {\bf 9},  310  (2013).

\bibitem{Berthier2014}
L. Berthier, Physical Review Letters {\bf 112},  220602  (2014).

\bibitem{Szamel2015}
G. Szamel, E. Flenner, and L. Berthier, Physical Review E {\bf 91},  062304  (2015).

\bibitem{DebetsTJCP2023}
V.~E. Debets and L.~M.~C. Janssen, The Journal of Chemical Physics {\bf 159},  14502  (2023).

\bibitem{Ni2013}
R. Ni, M.~A.~C. Stuart, and M. Dijkstra, Nature Communications {\bf 4},  2704  (2013).

\bibitem{Bi2016}
D. Bi, X. Yang, M.~C. Marchetti, and M.~L. Manning, Phys. Rev. X {\bf 6},  021011  (2016).

\bibitem{Liluashvili2017}
A. Liluashvili, J. {\'{O}}nody, and T. Voigtmann, Physical Review E {\bf 96},  062608  (2017).

\bibitem{Nandi2018}
S.~K. Nandi, R. Mandal, P.~J. Bhuyan, C. Dasgupta, M. Rao, and N.~S. Gov, Proceedings of the National Academy of Sciences {\bf 115},  7688  (2018).

\bibitem{Szamel2019}
G. Szamel, The Journal of Chemical Physics {\bf 150},  124901  (2019).

\bibitem{Mandal2020}
R. Mandal and P. Sollich, Physical Review Letters {\bf 125},  218001  (2020).

\bibitem{Reichert2021}
J. Reichert, S. Mandal, and T. Voigtmann, Physical Review E {\bf 104},  044608  (2021).

\bibitem{ReichertSoftMatter2021}
J. Reichert and T. Voigtmann, Soft Matter {\bf 17},  10492  (2021).

\bibitem{ReichertEPJE2021}
J. Reichert, L.~F. Granz, and T. Voigtmann, The European Physical Journal E {\bf 44},  27  (2021).

\bibitem{Debets2021}
V.~E. Debets, X.~M. de~Wit, and L.~M. Janssen, Physical Review Letters {\bf 127},  278002  (2021).

\bibitem{Debets2022}
V.~E. Debets and L.~M.~C. Janssen, Physical Review Research {\bf 4},  L042033  (2022).

\bibitem{Janzen2022}
G. Janzen and L.~M.~C. Janssen, Physical Review Research {\bf 4},  L012038  (2022).

\bibitem{Paoluzzi2022}
M. Paoluzzi, D. Levis, and I. Pagonabarraga, Communications Physics {\bf 5},  111  (2022).

\bibitem{Flenner2016}
E. Flenner, G. Szamel, and L. Berthier, Soft Matter {\bf 12},  7136  (2016).

\bibitem{Szamel2016}
G. Szamel, Physical Review E {\bf 93},  012603  (2016).

\bibitem{Feng2017}
M. Feng and Z. Hou, Soft Matter {\bf 13},  4464  (2017).

\bibitem{Berthier2017}
L. Berthier, E. Flenner, and G. Szamel, New Journal of Physics {\bf 19},  125006  (2017).

\bibitem{Flenner2020}
E. Flenner and G. Szamel, Physical Review E {\bf 102},  022607  (2020).

\bibitem{Debets2023}
V.~E. Debets, H. L{\"{o}}wen, and L.~M. Janssen, Physical Review Letters {\bf 130},  058201  (2023).

\bibitem{caprini2024self}
L. Caprini, B. Liebchen, and H. L{\"o}wen, Communications Physics {\bf 7},  153  (2024).

\bibitem{Supply2024}
See Supplemental Material at [publisher will insert URL] for description of simulation movies, analytic calculations, which include Refs. \cite{Chaikin1995, Henkes2020}.

\bibitem{Henkes2020}
S. Henkes, K. Kostanjevec, J.~M. Collinson, R. Sknepnek, and E. Bertin, Nature Communications {\bf 11},  1405  (2020).

\bibitem{Chen2010}
K. Chen, W.~G. Ellenbroek, Z. Zhang, D.~T. Chen, P.~J. Yunker, S. Henkes, C. Brito, O. Dauchot, W. Van~Saarloos, A.~J. Liu, and A.~G. Yodh, Physical Review Letters {\bf 105},  025501  (2010).

\bibitem{Henkes2012}
S. Henkes, C. Brito, and O. Dauchot, Soft Matter {\bf 8},  6092  (2012).

\bibitem{melio2024soft}
J. Melio, S.~E. Henkes, and D.~J. Kraft, Physical Review Letters {\bf 132},  078202  (2024).

\bibitem{Garcia2015}
S. Garcia, E. Hannezo, J. Elgeti, J.-F. Joanny, P. Silberzan, and N.~S. Gov, Proceedings of the National Academy of Sciences {\bf 112},  15314  (2015).

\bibitem{bialke2012crystallization}
J. Bialk{\'e}, T. Speck, and H. L{\"o}wen, Physical review letters {\bf 108},  168301  (2012).

\bibitem{fily2014freezing}
Y. Fily, S. Henkes, and M.~C. Marchetti, Soft matter {\bf 10},  2132  (2014).

\bibitem{Dash1999}
J.~G. Dash, Reviews of Modern Physics {\bf 71},  1737  (1999).

\bibitem{Gasser2009}
U. Gasser, Journal of Physics: Condensed Matter {\bf 21},  203101  (2009).

\bibitem{Li2020}
Y.-W. Li and M.~P. Ciamarra, Physical Review Letters {\bf 124},  218002  (2020).

\bibitem{digregorio2018full}
P. Digregorio, D. Levis, A. Suma, L.~F. Cugliandolo, G. Gonnella, and I. Pagonabarraga, Physical review letters {\bf 121},  098003  (2018).

\bibitem{Pasupalak2020}
A. Pasupalak, L. Yan-Wei, R. Ni, and M. Pica~Ciamarra, Soft Matter {\bf 16},  3914  (2020).

\bibitem{berthier2019glassy}
L. Berthier, E. Flenner, and G. Szamel, The Journal of Chemical Physics {\bf 150},  200901  (2019).

\bibitem{mandal2020extreme}
R. Mandal, P.~J. Bhuyan, P. Chaudhuri, C. Dasgupta, and M. Rao, Nature communications {\bf 11},  2581  (2020).

\bibitem{caprini2020hidden}
L. Caprini, U.~M.~B. Marconi, C. Maggi, M. Paoluzzi, and A. Puglisi, Physical Review Research {\bf 2},  023321  (2020).

\bibitem{keta24a}
Y.-E. Keta, J.~U. Klamser, R.~L. Jack, and L. Berthier, Phys. Rev. Lett. {\bf 132},  218301  (2024).

\bibitem{Hermans1952}
J. Hermans and R. Ullman, Physica {\bf 18},  951  (1952).

\bibitem{Shee2020ActiveMoments}
A. Shee, A. Dhar, and D. Chaudhuri, Soft Matter {\bf 16},  4776  (2020).

\bibitem{Chaudhuri2021}
D. Chaudhuri and A. Dhar, Journal of Statistical Mechanics: Theory and Experiment {\bf 2021},  013207  (2021).

\bibitem{Shee2022}
A. Shee and D. Chaudhuri, Journal of Statistical Mechanics: Theory and Experiment {\bf 2022},  013201  (2022).

\bibitem{Chaikin1995}
P.~M. Chaikin and T.~C. Lubensky, {\em {Principles of Condensed Matter Physics}} (Cambridge University Press, Cambridge, 1995).

\end{thebibliography}
\end{document}